\documentclass[letterpaper, oneside, 12pt]{article}
\usepackage{amsmath}
\usepackage{graphicx}%
\usepackage{amsfonts}%
\usepackage{amssymb}
\usepackage{xcolor}
\usepackage{overpic}
\usepackage[noend]{algpseudocode}
\usepackage{subfigure}
\usepackage{rotating}
\usepackage{epstopdf}
\usepackage{url}
\usepackage[linesnumbered, ruled, vlined]{algorithm2e}
\usepackage{comment}
\usepackage{lipsum}
\usepackage{enumitem}
\newlist{steps}{enumerate}{1}
\setlist[steps, 1]{label = Step \arabic*:}
\usepackage{mathtools}
\usepackage{multirow}
\usepackage{pgfplots}
\usepackage{natbib}
\usepackage{pst-solides3d}
\usepackage{booktabs}
\usepackage{tabularray}
\usepackage{multirow}
\usepackage[section]{placeins}
\usepackage{subcaption}
\usepackage{newfloat}
\usepackage{caption}
\usepackage{lscape}

\usepackage{setspace}
\usepackage[hmargin=1in,vmargin=1in]{geometry}

\usepackage[compact]{titlesec}
\titlespacing{\section}{0pt}{*0.8}{*0.8}
\titlespacing{\subsection}{0pt}{*0.8}{*0.8}
\titlespacing{\subsubsection}{0pt}{*0.8}{*0.8}

\usepackage{hyperref}
\newcommand{\ba}{ {\boldsymbol a} }

\newcommand{\bb}{ {\boldsymbol b} }

\newcommand{\bD}{ {\boldsymbol D} }

\newcommand{\bg}{ {\boldsymbol g} }

\newcommand{\bh}{ {\boldsymbol h} }

\newcommand{\bI}{ {\boldsymbol I} }

\newcommand{\bK}{ {\boldsymbol K} }

\newcommand{\bs}{ {\boldsymbol s} }

\newcommand{\bu}{ {\boldsymbol u} }

\newcommand{\bv}{ {\boldsymbol v} }

\newcommand{\bw}{ {\boldsymbol w} }
\newcommand{\bW}{ {\boldsymbol W} }
\newcommand{\bx}{ {\boldsymbol x} }

\newcommand{\by}{ {\boldsymbol y} }

\newcommand{\bz}{ {\boldsymbol z} }
\newcommand{\bZ}{ {\boldsymbol Z} }

\newcommand{\bbeta}{ {\boldsymbol \beta} }

\newcommand{\bDelta}{ {\boldsymbol \Delta} }

\newcommand{\bepsilon}{ {\boldsymbol \epsilon} }

\newcommand{\bmu}{ {\boldsymbol \mu} }

\newcommand{\bet}{ {\boldsymbol \eta} }

\newcommand{\btheta}{ {\boldsymbol \theta} }

\newcommand{\bPsi}{ {\boldsymbol \Psi} }

\newcommand{\bzero}{ {\boldsymbol 0} }

\newcommand{\qed}{\nobreak \ifvmode \relax \else
      \ifdim\lastskip<1.5em \hskip-\lastskip
      \hskip1.5em plus0em minus0.5em \fi \nobreak
      \vrule height0.75em width0.5em depth0.25em\fi}

\date{} 

\title{Deep Generative Modeling with Spatial and Network Images: An Explainable AI (XAI) Approach}

\author{
Yeseul Jeon$^{1,2}$, Rajarshi Guhaniyogi$^{1,\dagger}$, Aaron Wolfe Scheffler$^{2, \dagger}$\\
\small
$^{1}$Department of Statistics, Texas A\&M University, College Station, TX, USA\\
\small
$^{2}$Department of Epidemiology \& Biostatistics, University of California San Francisco, San Francisco, CA, USA.\\
\small
$\dagger$ These authors have contributed equally to this work.
}

\begin{document}

\maketitle
\begin{abstract}
This article addresses the challenge of modeling the amplitude of spatially indexed low frequency fluctuations (ALFF) in resting state functional MRI as a function of cortical structural features and a multi-task coactivation network in the Adolescent Brain Cognitive Development (ABCD)
Study. It proposes a generative model that integrates effects of spatially-varying inputs and a network-valued input using deep neural networks to capture complex non-linear and spatial associations with the outcome. The method models spatial smoothness, accounts for subject heterogeneity and complex associations between network and spatial images at different scales, enables accurate inference of each images effect on the outcome image, and allows prediction and inference with uncertainty quantification via Monte Carlo dropout, contributing to one of the first Explainable AI (XAI) frameworks for image-on-image regression with spatial and network images. The model is highly scalable to high-resolution data without the heavy pre-processing or summarization often required by Bayesian methods. We applied the XAI model to the ABCD data which revealed associations between cortical features and ALFF throughout the entire brain. Our model performed comparably or a bit better than existing methods in predictive accuracy but provided superior uncertainty quantification and faster computation, demonstrating its effectiveness for large-scale neuroimaging analysis. A supplementary file contains additional simulation results and additional results from the ABCD data analysis.
\end{abstract}

\section{Introduction}\label{sec:intro}

Recently, a critical concern in the field of neuroimaging has been unraveling the intricate relationships between images that capture different aspects of brain structures and function using magnetic resonance imaging (MRI). This involves integrating multimodal imaging data—specifically, (a) \emph{brain network information} from functional MRI (fMRI), which quantifies connectivity between nodes, and (b) \emph{brain spatial information} from structural MRI (sMRI) and fMRI, which offers ROI-level information. These data are hierarchically structured, with ROIs nested within network nodes, offering insights at multiple scales.

For example, this article is motivated by the challenge of modeling the amplitude of spatially indexed low-frequency fluctuations (ALFF) in resting state functional MRI (rs-fMRI) as a function of cortical structural features measured via sMRI and a multi-task coactivation network capturing coordinated patterns of brain activation measured via task-based fMRI (t-fMRI) in a large neuroimaging study of adolescents. Recent studies in cognitive neuroscience suggest that ALFF serves as an indicator of brain functional integrity and supports information processing \citep{Fu2017}. It has been linked to atypical cognitive development, cognitive decline, and various psychiatric disorders, highlighting its relevance for both basic and clinical neuroscience research \citep{Sheng2021}. What remains unclear from current research is how spatial and network features of the brain contribute jointly to explain variability in ALFF. A major challenge in bridging this gap is a lack of available methods that can accommodate heterogeneity in multimodal imaging data and the immense data volume produced in large neuroimaging studies while simultaneously providing principled inference.

The primary scientific objectives of this study encompass whole-brain analysis and are threefold: (a) to predict the spatially indexed ALFF outcome using both cortical structural predictors and a brain network predictor;
(b) to jointly model the spatially varying effects of the structural predictors on ALFF through interpretable regression functions, enabling inference on the contributions of cortical features; and (c) to quantify uncertainty in model components and predictions, facilitating principled statistical inference. These scientific goals must be met while accommodating the data volume produced from large neuroimaging studies required to detect the often-subtle effects that link brain image features. 

To achieve these goals, this article proposes a two-stage generative model for input and outcome images. In the first stage, each edge of the network predictor is modeled as an interaction between latent vectors associated with the nodes, enabling the estimation of node-specific latent effects. The second stage adopts a semi-parametric regression approach that incorporates additive contributions from both spatial and network predictors. Spatial effects are captured using nonlinear spatially varying coefficient functions, while the influence of each network node is modeled as a function of the latent effects estimated in the first stage.

These nonlinear functions are modeled using deep neural networks (DNNs) with a dropout architecture. To perform inference, we employ \emph{Monte Carlo dropout} (MC dropout) architecture \citep{gal2016dropout}, leveraging its connection to deep Gaussian processes (GP). This allows us to predict the output image and to estimate the regression effects of spatial predictors and nonlinear node effects, all with quantified uncertainty, offering one of the first \emph{explainable AI} methods in the context of heterogeneous image-on-image regression.

\subsection{Novelty of the Proposed Approach}
\textbf{(1) Integration of multimodal data for image prediction.} The proposed framework seamlessly integrates spatial and network images across multiple scales, leveraging their structural relationships through a flexible, nonlinear, multiscale modeling approach to enable predictive inference for an outcome image. Image-on-image regression exploiting the structures of both spatial and network inputs and their nonlinear effects jointly on spatial outputs remains largely underexplored, making this one of the first principled methods to address the problem.
\textbf{(2) Explainable inference.} Our approach leverages the connection between deep Gaussian processes and
dropout-based deep neural networks (DNNs) to generate approximate posterior samples of DNN parameters. This approach leads to both point estimation and uncertainty quantification for the effects of spatial predictors and network nodes, as well as for the predicted outcome image. This enables interpretable inference and positions our method as one of the \emph{first deep learning-based} frameworks to offer explainability in image-on-image regression. \textbf{(3) Computational efficiency.} The model is highly scalable, with computation growing linearly in both the sample size and the number of ROIs, enabling its future use in larger neuroimaging studies. \textbf{(4) State-of-the-art performance.} Empirical results demonstrate that our approach achieves competitive point estimates and predictions while substantially outperforming existing methods in uncertainty quantification.

\subsection{Related Work}
Image datasets represent high-dimensional, structured data characterized by complex dependency patterns. To model these intricacies, many studies have leveraged deep learning architectures capable of capturing nonlinear dependencies through multiple layers of parameterization \citep{tsuneki2022deep, razzak2018deep}. Bayesian deep learning frameworks are especially attractive for their flexibility of offering coherent probabilistic inference. Bayesian methods are increasingly applied in medical imaging for tasks like classification, segmentation, and tumor growth prediction \citep{zou2023review, prince2023uncertainty, jeon2024bayesian}. However, deep learning approaches for predictive inference of outcome images based on input images, popularly referred to as the image-on-image regression, remain notably limited.

Several recent studies align with our framework by incorporating spatial information into Bayesian deep learning \citep{kirkwood2022bayesian, li2023semiparametric, zammit2023spatial, mateu2022spatial}. For example, \citet{mateu2022spatial} used variational autoencoder-based generative networks for spatio-temporal point processes, while \citet{kirkwood2022bayesian} applied Bayesian deep neural networks to spatial interpolation. \citet{li2023semiparametric} proposed replacing splines or kernels with DNNs to model nonlinear spatial variation in semiparametric regression. Spatial Bayesian neural networks (SBNNs) \citep{zammit2023spatial} enhance flexibility by incorporating spatial embeddings and spatially varying parameters within BNNs, though at higher computational cost. Unlike our method, which directly models spatially varying coefficients, SBNNs rely on spatial basis functions and learn weight and bias parameters in neural networks via embedding layers.

For interpretable machine learning in the context of image-on-image regression, one of the most relevant methods is BIRD-GP \citep{ma2023bayesian}—a two-stage Bayesian model that employs a combination of GPs and DNNs. Specifically, this approach uses neural networks to learn basis expansions of input and output images, followed by nonlinear regression on vectorized projections of the input and output images. While BIRD-GP provides predictive uncertainty through the posterior distribution, its two-stage design may lead to increased computational cost and a greater risk of overfitting. The proposed approach offers competitive inference to BIRD-GP in all empirical investigation and is much faster in computation.

The remainder of the paper is structured as follows. Section 2 describes the ABCD neuroimaging dataset and outlines the associated scientific objectives. Section 3 presents the proposed methodology and explains how the framework enables interpretable inference using DNNs. Section 5 assesses performance through simulation studies, followed by an application to the ABCD data in Section 6. Section 7 concludes with a discussion. Additional results on simulations and ABCD data can be found in the supplementary file.

\section{Imaging Data from the ABCD Study}\label{sec:data_description}

This article explores a clinical application involving multimodal imaging data from 
Adolescent Brain Cognitive Development (ABCD) Study \citep{casey2018}. The primary aim is to investigate how spatial patterns in variability of rs-fMRI signals across time, reflecting the amplitude of low frequency oscillations, relates to both spatial brain characteristics and a multi-task brain coactivation network. We begin by introducing our motivating scientific question. 

\subsection{Motivating Scientific Question}

fMRI captures blood-oxygen-level dependent (BOLD) signals in the brain while subjects are at rest or engaged in tasks, enabling the assessment of localized neuronal activity. Fluctuations in the BOLD signal across time—typically quantified through variance or standard deviation of the time series—are now increasingly viewed as meaningful indicators of brain function rather than random noise, and are interpreted as reflecting the amplitude of low-frequency fluctuations (ALFF) defined as the power of the BOLD signal between $0.01$ and $0.10$ Hz \citep{Zuo2010,hagler2019, Wang2021}. Existing studies have linked ALFF with structural properties of the brain via diffusion MRI (dMRI) \citep{Wang2021}; gray and white matter content assessed through structural MRI (sMRI) \citep{Zuo2010}), patterns of functional connectivity in brain networks (rs-fMRI \citep{Fu2017, Sheng2021, Tomasi2016}), and response to task performance (task-based fMRI (t-fMRI) \citep{tomasi2019}). 

Building on this evidence, emerging findings in both computational and cognitive neuroscience suggest that ALFF serves as a marker of functional brain integrity, supports core processes of long distance information handling, and changes in response to task-activation \citep{Zuo2010, tomasi2019}. To this end, the potential role of ALFF in supporting human brain networks can be assessed via associations with resting state networks and task-based brain activation \citep{cole2016, ye2022}. What is unclear from current work is how structural and network characteristics of the brain jointly contribute to variation in observed ALFF. Specifically, we are interested in modeling ALFF as a function of brain structure (cortical thickness and gray-white matter intensity contrasts measured via sMRI) and a multi-task coactivation network capturing coordinated patterns of brain activation during the execution of several cognitive tasks (measured via t-fMRI). In modeling variation in ALFF from multiple images, it is desirable to provide interpretable mappings of each input image onto the outcome as well as principled uncertainty quantification in terms of model components and prediction. 

Most statistical modeling of ALFF to date has concentrated on linking BOLD signal variability or ALFF distributions to either structural or functional brain characteristics by extracting targeted predictor variables from a \emph{single} imaging modality and applying relatively straightforward statistical techniques. These investigations have included t-test comparisons of ALFF across regions dominated by gray versus white matter \citep{Zuo2010}, regional modeling of ALFF as a function of global white matter maturation through linear mixed effects models \citep{Wang2021}, linear regression modeling to associate ALFF with task-based activation maps \citep{tomasi2019}, and studies connecting ALFF to resting-state connectivity using parametric mapping and correlational analyses of various functional graph metrics \citep{Sheng2021, Tomasi2016}. Recently, source localization has been used to associate low-frequency fluctuations captured by magnetoencephalography with ALFF derived from fMRI \citep{Zhang2023}. However, no study has jointly modeled ALFF using both structural and network images, despite growing evidence that these modalities provide complementary information \citep{calhoun2012}.

\begin{figure}[h!]
    \centering
    \includegraphics[scale=0.70]{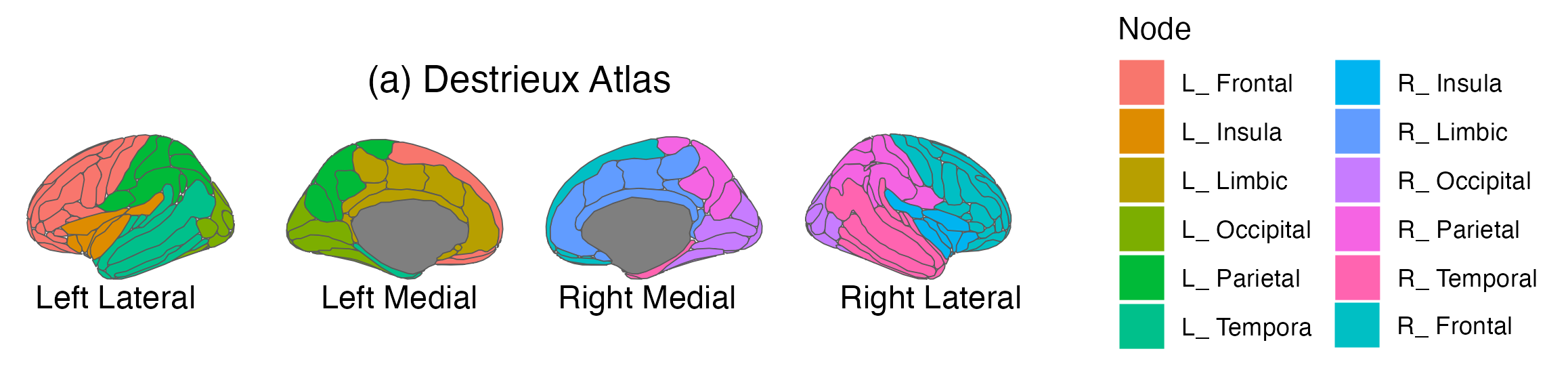}\\
    \includegraphics[scale=0.30]{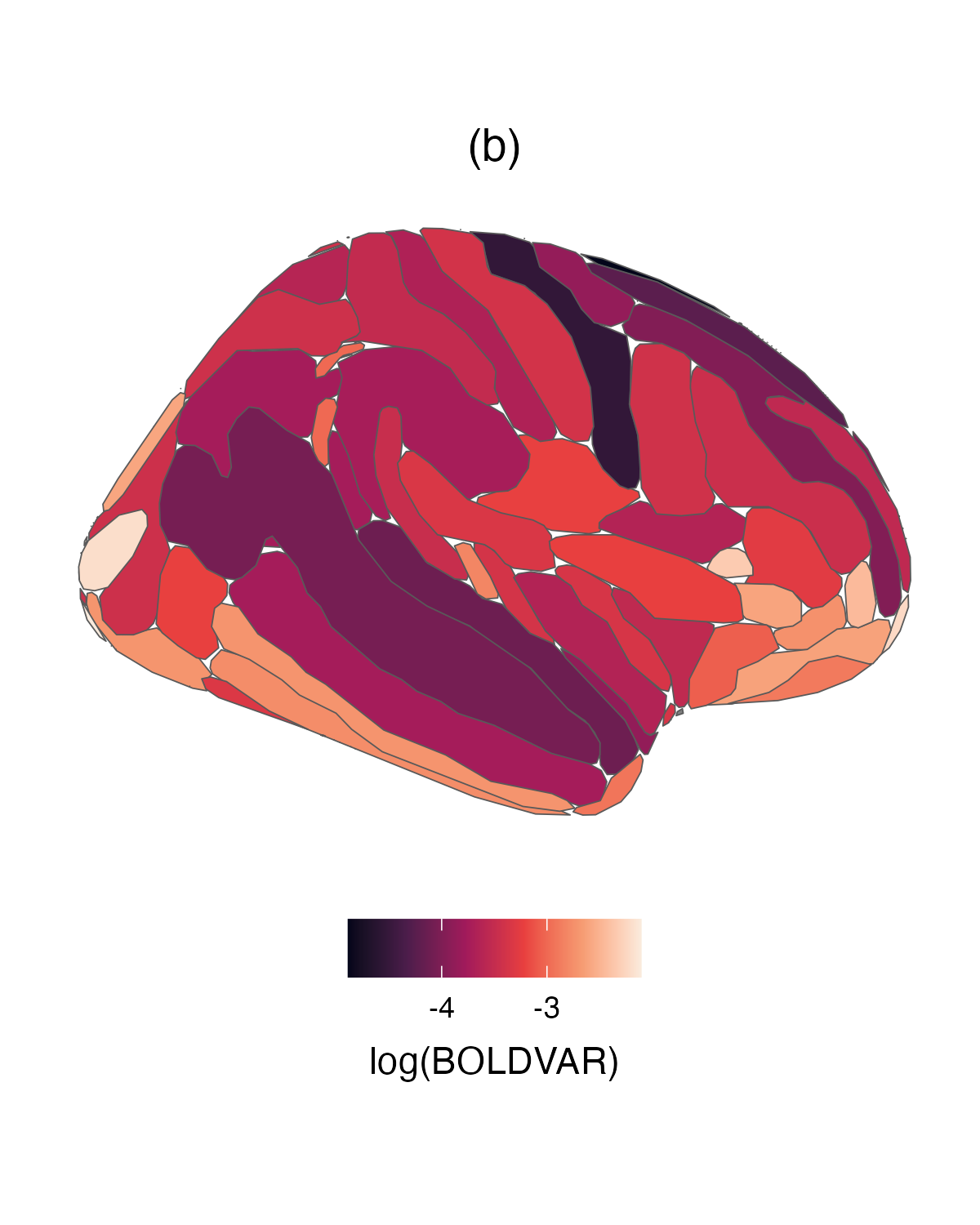}
    \includegraphics[scale=0.30]{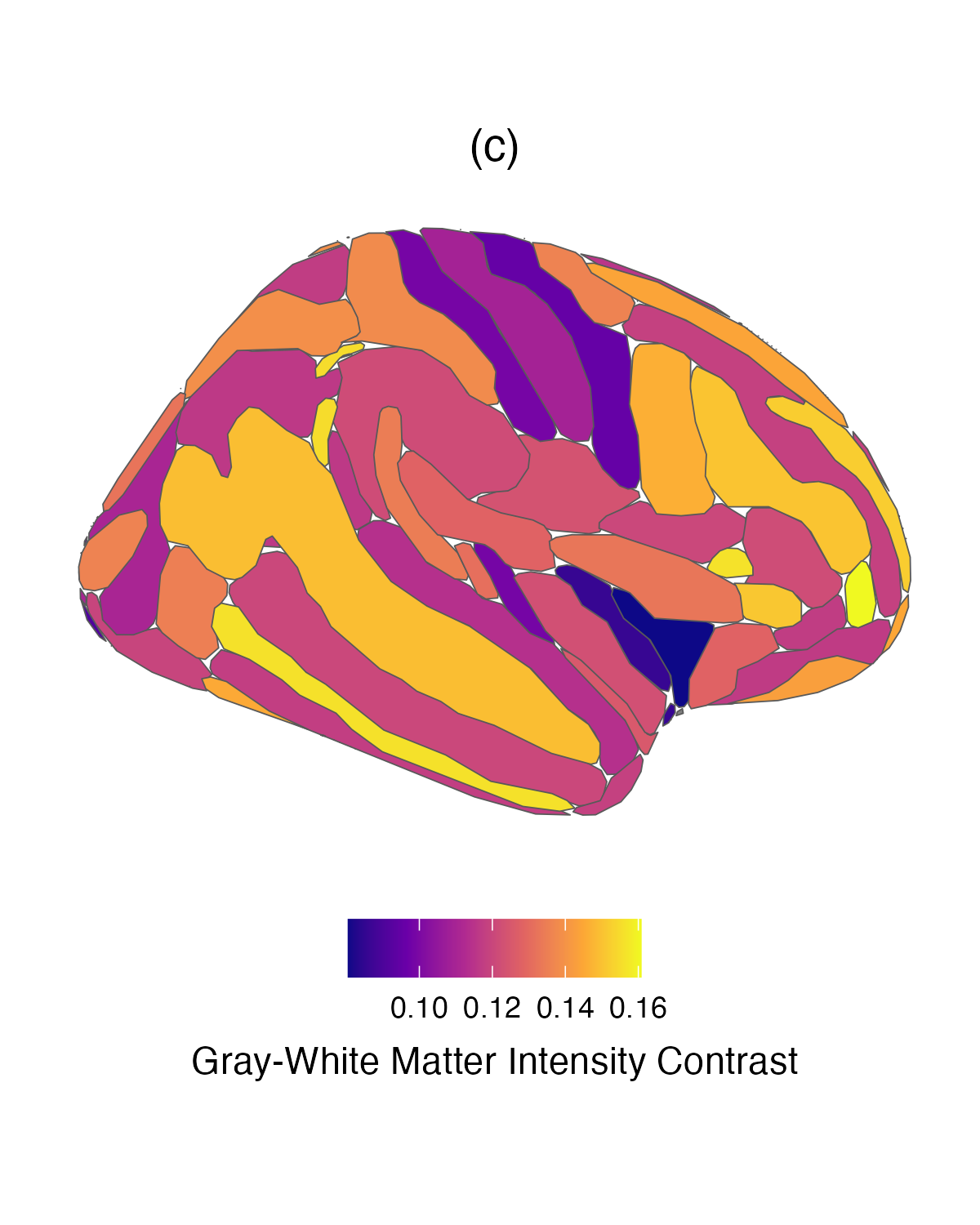}
         \includegraphics[scale=0.30]{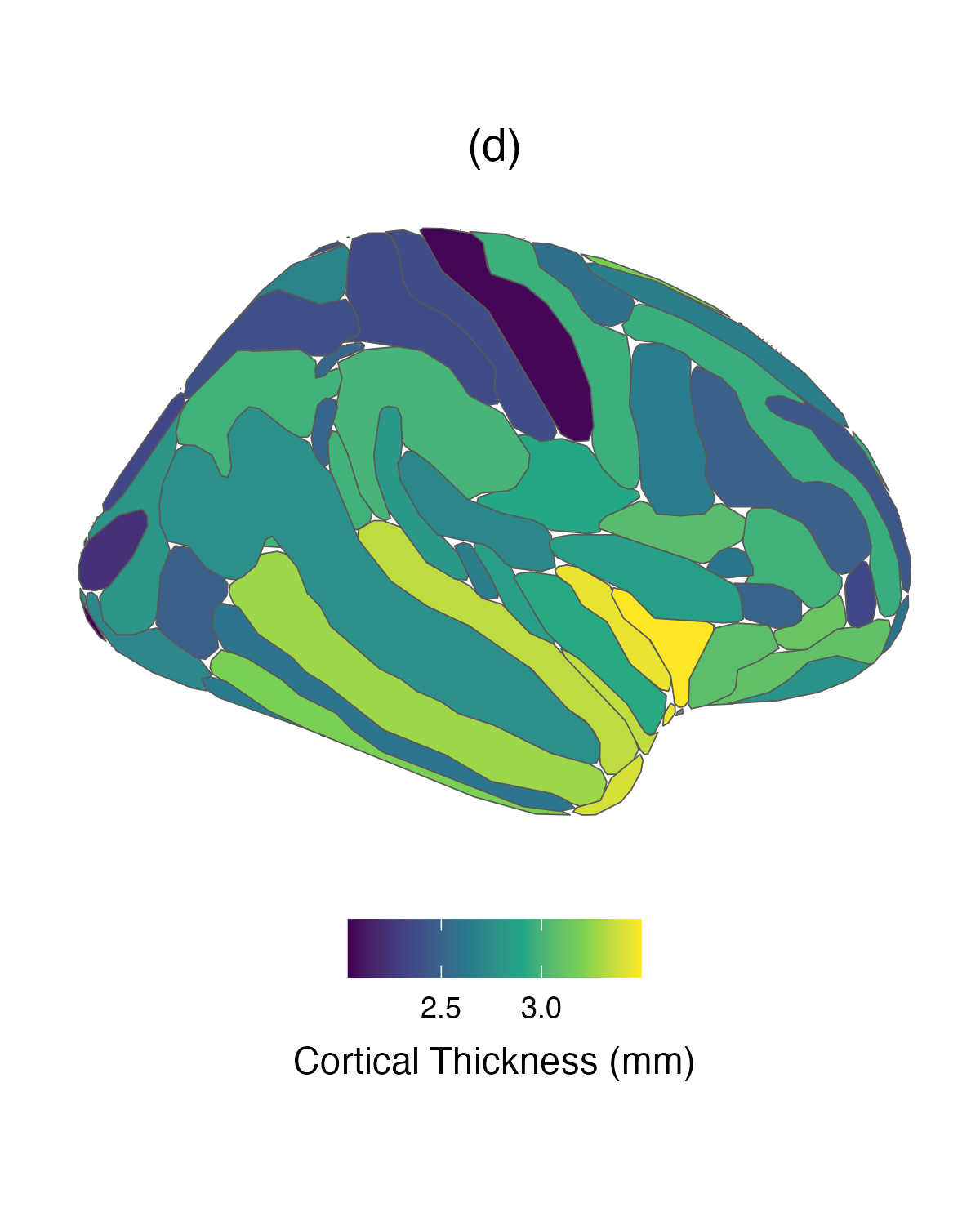}
    \includegraphics[scale=0.30]{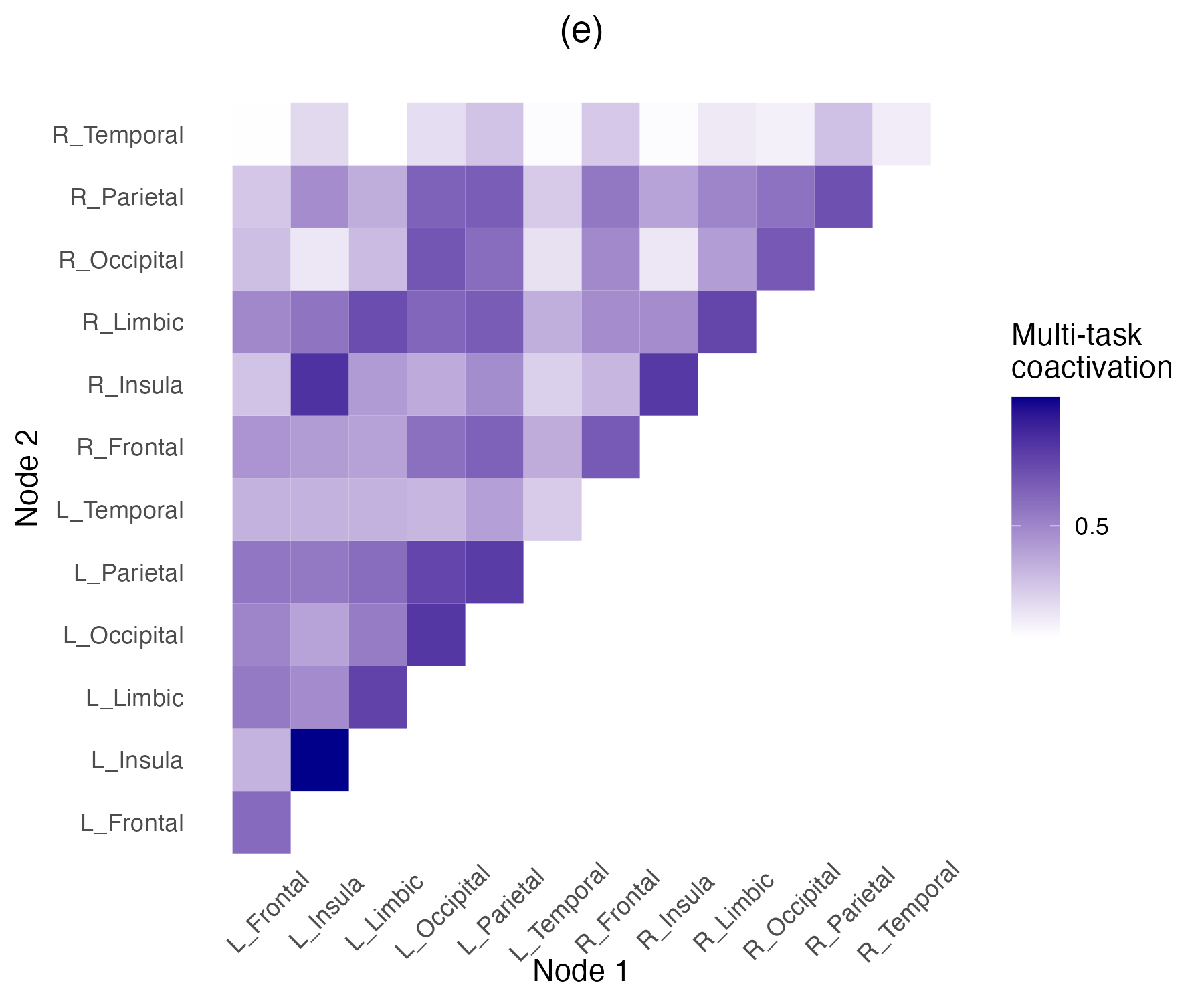}
    \caption{(a) The Destriuex atlas used to parcellate our motivating ABCD imaging data into 148 ROIs grouped by nodes defined by hemisphere (left, right) and lobes (frontal, parietal, occipital, insula, and limbic). (b) Outcome of interest: variance in the rs-fMRI BOLD time series (BOLDVAR) which is a proxy for amplitude of low frequency fluctuations (ALFF). (c) Structural predictor: cortical gray/white matter intensity contrast (GWMIC). (d) Structural predictor: cortical thickness (CT). (e) Network predictor: multi-task coactivation matrix representing brain coactivation across $12$ nodes in response to an array of cognitive tasks.}
\label{fig:data}
\end{figure}

\subsection{Clinical Imaging}
We used data from the ABCD Study 5.0 Tabulated Release (http://dx.doi.org/10.15154/8873-zj65), focusing on baseline imaging measures from children aged 9–10 years. Imaging data were acquired and processed as outlined in \cite{hagler2019}. All images were aligned to the Montreal Neurological Institute (MNI) template space and parcellated to the Destrieux atlas (Figure \ref{fig:data}(a); \citep{destriuex2010}) yielding 148 spatially defined regions of interests (ROIs) grouped into 12 nodes defined by brain hemisphere (left, right) and lobe (frontal, parietal, occipital, insula, and limbic). Thus, the data has hierarchical structure in which ROIs are organized within a collection of mutually exclusive nodes. Our outcome measure is the variance of the BOLD signal measured via rs-fMRI (BOLDVAR; Figure \ref{fig:data}(b)) at each ROI which is considered a proxy for ALFF by the ABCD study team \citep{hagler2019} and is log transformed to better satisfy model assumptions in Section \ref{sec:methods}. We consider two structural cortical predictors measured via sMRI at each ROI: cortical gray and white matter intensity contrasts (GWMIC; Figure \ref{fig:data}(c)), thought to capture brain development, and cortical thickness (CT; Figure \ref{fig:data}(d)), a local measure of gray matter thickness. We also consider a network image predictor which captures multi-task brain coactivation (Figure \ref{fig:data}(e)), a measure of coordinated neuronal activation across the brain while performing different tasks. Specifically, we obtained task-activation maps measuring local BOLD signals via t-fMRI collected on subjects while engaged in three tasks (money incentive delay task, stop signal task, and emotional version of the n-back task) with four contrasts per task yielding twelve task-activation maps as described in \cite{ye2022}. Based on the twelve task-activation maps, we constructed a correlation matrix among ROIs in the brain for each subject which we then averaged by node to yield a $12 \times 12$ multi-task coactivation matrix representing patterns of brain coactivation across hemispheres and lobes in response to a diverse array of tasks. 
Our aim is to expand the investigation of ALFF by integrating multimodal image inputs to better understand how structural and network-level factors together influence local variability in rs-fMRI signals. To address this, we introduce novel regression framework, as outlined in Section~\ref{sec:methods}.

\section{Methodology}\label{sec:methods}
\subsection{Notations}
The dataset consists of $n$ observations, each containing input and outcome images. For $i=1,..,n$, let $\bZ_i$ denotes an undirected brain network, represented by a symmetric $V\times V$ matrix. Each entry $z_{i,(v,v')}$ quantifies the connectivity strength between functional nodes $\mathcal{G}_v$ and $\mathcal{G}_{v'}$. These networks are undirected and contain no self-loops, so $z_{i,(v,v')}=z_{i,(v',v)}$, and $z_{i,(v,v)}=0$. The $V$ functional nodes are together denoted by $\mathcal{G}=\{\mathcal{G}_1,...,\mathcal{G}_V\}$. The $v$th node $\mathcal{G}_v$ contains $J_v$ ROIs, such that the total number of ROIs is $\sum_{v=1}^V J_v=J$. Let $y_{i}(\bs_{v,j})\in\mathbb{R}$ denote the spatial outcome, and $\bx_{i}(\bs_{v,j})\in\mathbb{R}^Q$ represent spatial inputs at the $j$th ROI of the $v$th functional node, with spatial co-ordinate $\bs_{v,j}\in\mathbb{R}^d$, for the $i$th subject. 
The notation $\ba \odot \bb$ stands for the Hadamard product describing the elementwise product between two vectors $\ba$ and $\bb$. In our context, $y_i(\bs_{v,j})$ is the log-transformed BOLDVAR, $\bx_i(\bs_{v,j})$ represents the vector of GWMIC and CT, while $\bZ_i$ is the coactivation network matrix.

\subsection{Proposed Explainable Artificial Intelligence (XAI) Model}
\noindent We propose a two-stage framework: first, we estimate subject-specific overall network connectivity and nodal effects for the $V$ functional nodes from brain networks of subjects, capturing inter-nodal associations in a latent space. Second, we develop an explainable AI framework to predict the outcome image from spatial input images and estimated nodal effects.

\subsubsection{First Stage: Estimate Nodal Effects of the Network Predictor}
The first stage of our framework fits a parsimonious probabilistic model to the edges of the undirected network predictor $\bZ_i$, for subjects $i=1,\cdots,n$. We represent node interactions within a lower-dimensional latent space that captures the network structure without imposing overly restrictive assumptions. Specifically, we model the edge weights as:
\begin{align}\label{eq:first_stage}
g^{-1}(E[z_{i,(v,v')}|\eta_i,\bu_{i,v},\bu_{i,v'},\sigma_i^2])=\eta_i-||\bu_{i,v}-\bu_{i,v'}||+\epsilon_{i,(v,v')},\:\:\epsilon_{i,(v,v')}\sim N(0,\sigma_i^2),    
\end{align}
where $g(\cdot)$ is a link function tailored to the type of edge weights (binary, categorical, or continuous). The parameter $\eta_i$ captures overall degree of subject-specific network connectivity, while $\sigma_i^2$ denotes edge-level variability for the $i$th subject. The vector $\bu_{i,v}=(u_{i,v,1},...,u_{i,v,R})^T\in\mathbb{R}^R$ is a latent coordinate for functional node $\mathcal{G}_v$ for the $i$th subject, encoding its position in an $R$-dimensional latent space. The strength of interaction between nodes $\mathcal{G}_v$ and $\mathcal{G}_{v'}$ for subject $i$ is modeled via the Euclidean distance between their latent positions $\mathbf{u}_{i,v}$ and $\mathbf{u}_{i,v'}$—smaller distances indicate stronger interaction. The parameter $R$ is the intrinsic latent dimension, and may be selected through cross-validation or more adaptive strategies \citep{guha2020bayesian} at a greater computational cost. As model performance is generally stable across a range of $R$, we opt for fixed $R$ to maintain efficiency.

We assign a prior on each $\eta_i{\sim}N(0,1)$ and the priors $\bu_{i,v}\stackrel{i.i.d.}{\sim} N(\bzero,\bI_R)$, $\sigma_i^2\sim IG(\alpha,\beta)$, estimating the joint posterior distribution using Markov Chain Monte Carlo (MCMC) implemented in the \texttt{Nimble} software \citep{nimble}. Since the model (\ref{eq:first_stage}) depends on the latent effects $\bu_{i,v}$'s only through pairwise distances, it is invariant to the rotation of latent effects. We align estimates across MCMC samples to a common orientation by applying a ``Procrustean transformation" \citep{borg2005modern}. Upon the transformation, the latent effects are \emph{comparable} across subjects and across nodes within a subject. Posterior means of the aligned latent nodal estimate $\widehat{\bu}_{i,v}$ and posterior means for the overall degree of network connectivity $\widehat{\eta}_i$ are used as subject-level inputs in the second-stage regression model.

\subsubsection{Second Stage: Additive Non-linear Regression Function Estimation Through Deep Neural Networks}
In the second stage of our approach, we formulate a non-linear regression function to delineate association between the spatial outcome $y_{i}(\bs_{v,j})$ and spatial inputs $\bx_{i}(\bs_{v,j})$, the estimated overall connectivity $\widehat{\eta}_i$ and the estimated nodal effects $\widehat{\bu}_{i,v}$ corresponding to the $v$th functional node. Let the features from the network input be denoted by $\bg_{i,v}=(\widehat{\bu}_{i,v},\widehat{\eta}_i)^T$ More precisely, our modeling approach involves
\begin{align}\label{additive_reg}
y_{i}(\bs_{v,j})=\bx_{i}(\bs_{v,j})^T\bbeta(\bs_{v,j})+h(\bg_{i,v})+\epsilon_{i}(\bs_{v,j}),\:\:j=1,...,J_v;\:v=1,...,V,
\end{align}
where the idiosyncratic errors are assumed to follow $\epsilon_{i}(\bs_{v,j})\stackrel{i.i.d.}{\sim} N(0,\tau^2)$. Note that the data are mean-standardized.
Here $h(\cdot)$ encodes the non-linear relationship between the outcome image at the $j$th ROI with spatial location $\bs_{v,j}$ within the $v$th functional node, and $\widehat{\bu}_{i,v}$ and $\widehat{\eta}_i$. Although $\widehat{\bu}_{i,v}$ and $\widehat{\eta}_i$ are the latent features used as surrogates for the network predictor, the downstream analysis produces stable predictions with these surrogates. The regression effects of spatial predictors $\bx_{i}(\bs_{v,j})$ is encoded in the coefficient vector $\bbeta(\bs_{v,j})=(\beta_{1}(\bs_{v,j}),...,\beta_{Q}(\bs_{v,j}))^T\in\mathbb{R}^Q$ which also varies over spatial locations, and are referred to as varying-coefficients (VCs). 
VCs combine the flexibility of non-parametric models to capture sufficiently complex dependencies between the outcome and predictors with the interpretability of parametric models \citep{gelfand2003spatial, guhaniyogi2022distributed}. 

Traditionally, $\bbeta(\cdot)$ and $h(\cdot)$ are modeled with Gaussian process (GP) priors, but their estimation is computationally intensive, scaling as $O(J^3)$ and $O(n^3)$. Basis function approximations improve efficiency \citep{morris2015functional, reiss2017methods} but often oversmooth. Recent scalable GP methods \citep{guhaniyogi2022distributed} enhance estimation for large data but still fall short in delivering fast, uncertainty-aware inference needed for modern imaging studies with thousands of subjects and regions.

There is a growing literature on standard deep neural networks (DNNs) to train dependency structures within spatial data~\citep{wikle2023statistical}, which is ideal for reducing computational complexity along with capturing spatial dependence. Expanding upon that literature, for $q=1,..,Q$, we model \eqref{additive_reg} using DNN with dropout vectors applied to weight matrices and bias vectors, given by,
\begin{align}\label{eq:XAI}
    &\beta_{q}(\bs_{v,j}) = \sigma^{(\beta)}_{q,L_\beta} \Big(\mathbf{W}^{(\beta)}_{q,L_\beta} \sigma^{(\beta)}_{q,L_\beta-1} \Big( \cdots \sigma^{(\beta)}_{q,1} \Big(\mathbf{W}^{(\beta)}_{q,1}\bs_{v,j} +\mathbf{b}^{(\beta)}_{q,1}\Big) \odot \bz^{(\beta)}_{q,1} \cdots \Big)\odot \bz_{q,L_{\beta}-1}^{(\beta)}+\mathbf{b}^{(\beta)}_{q,L_\beta} \Big) \nonumber\\
    & h(\bg_{i,v}) = \sigma_{L_h}^{(h)} \Big(\mathbf{W}_{L_h}^{(h)} \sigma_{L_h-1}^{(h)} \Big( \cdots  \sigma_{1}^{(h)} \Big(\mathbf{W}_{1}^{(h)}\bg_{i,v} +\mathbf{b}_{1}^{(h)}\Big)\odot\bz_{1}^{(h)} \cdots \Big)\odot \bz_{L_{h}-1}^{(h)} +\mathbf{b}_{L_h}^{(h)} \Big), 
\end{align}
where $\odot$ denotes the element-wise product between two vectors/matrices of the same dimensions, $\sigma^{(\beta)}_{q,l_1}(\cdot)$ is an activation function of the $l_1$th layer for the DNN construction of the coefficient function $\beta_{q}(\cdot)$, $l_1=1,\cdots,L_{\beta}$, and $\sigma_{l_2}^{(h)}(\cdot)$ is an activation function of the $l_2$th layer for the DNN representation of the function $h(\cdot)$, $l_2=1,\cdots,L_h$. Assume that the $l_1$th layer of the DNN construction of $\beta_{q}(\cdot)$ and $l_2$th layer of DNN construction of $h(\cdot)$ include $k_{l_1}^{(\beta)}$ and $k_{l_2}^{(h)}$ neurons, respectively. The matrix $\bW_{q,l_1}^{(\beta)}\in\mathbb{R}^{k_{l_1}^{(\beta)}\times k_{l_1-1}^{(\beta)}}$ is the weight parameters connecting the $l_1$ and $(l_1-1)$th layer for the DNN construction of $\beta_{q}(\cdot)$, and $\bW_{l_2}^{(h)}\in\mathbb{R}^{k_{l_2}^{(h)}\times k_{l_2-1}^{(h)}}$ is the corresponding matrix  connecting the $l_2$ and $(l_2-1)$th layer for the DNN construction $h(\cdot)$, respectively. The $k_{l_1}^{(\beta)}$ and  $k_{l_2}^{(h)}$ dimensional vectors $\bb_{q,l_1}^{(\beta)}$ and $\bb_{l_2}^{(h)}$ are the bias parameters in the construction of DNN architectures for $\beta_q(\cdot)$ and $h(\cdot)$, respectively. The vectors $\bz_{q,l_1}^{(\beta)} \in \mathbb{R}^{k^{(\beta)}_{l_1-1}}$ and $\bz_{l_2}^{(h)} \in \mathbb{R}^{k^{(h)}_{l_2-1}}$ have binary entries in $\{0,1\}$, referred to as the \emph{dropout vectors}, which randomly drop nodes in different layers of a neural network.  

A traditional DNN approach proceeds by training the model by minimizing the following loss function with regularization terms as
\begin{align}\label{eq:loss}
&\mathcal{L}_{\text{DNN}}(\btheta,\tau^2) = \frac{1}{2N} \sum_{i=1}^{n} \sum_{v=1}^V\sum_{j=1}^{J_v}\frac{\left( y_{i}(\bs_{v,j}) - \widehat{y}_{i}(\bs_{v,j}) \right)^2}{\tau^2} 
+ \sum_{q=1}^{Q} \sum_{l_1=1}^{L_\beta} \lambda_{l_1}^{(W, \beta_q)} \left\| \mathbf{W}_{q,l_1}^{(\beta)} \right\|_2^2 
\nonumber \\
&\qquad\qquad\qquad\qquad + \sum_{q=1}^{Q} \sum_{l_1=1}^{L_\beta} \lambda_{l_1}^{(b, \beta_q)} \left\| \mathbf{b}_{q,l_1}^{(\beta)} \right\|_2^2  + \sum_{l_2=1}^{L_h} \lambda_{l_2}^{(W, h)} \left\| \mathbf{W}_{l_2}^{(h)} \right\|_2^2 
+ \sum_{l_2=1}^{L_h} \lambda_{l_2}^{(b, h)} \left\| \mathbf{b}_{l_2}^{(h)} \right\|_2^2,
\end{align}
where $N=n\times J$, $J=\sum_{v=1}^V J_v$, and $\widehat{y}_{i}(\bs_{v,j}) = \bx_{i}(\bs_{v,j})^T\bbeta(\bs_{v,j}) + h(\bg_{i,v})$ is the point prediction from the model. Here $\btheta=\{(\bW_{q,l_1}^{(\beta)},\bb_{q,l_1}^{(\beta)}): q=1,..,Q; l_1=1,..,L_\beta\}\cup \{(\bW_{l_2}^{(h)},\bb_{l_2}^{(h)}): l_2=1,..,L_h\}$ denotes the weights and bias parameters for the DNN architecture of $\beta_q(\cdot)$ and $h(\cdot)$. The parameters $\{(\lambda_{l_1}^{(W, \beta_q)}, \lambda_{l_1}^{(b, \beta_q)}): l_1=1,..,L_\beta; q=1,..,Q\}$ and $\{(\lambda_{l_2}^{(W, h)},\lambda_{l_2}^{(b, h)}): l_2=1,...,L_h\}$ are penalty parameters controlling shrinkage for weight and bias parameters. Equation~\eqref{eq:loss} imposes regularization on the connections between layers, minimizing the contribution of interconnections between many pairs of neurons in two consecutive layers, preventing DNN from overfitting.
The aforementioned DNN framework offers a data-driven optimal point estimate $\widehat{\btheta}$ for the parameters $\btheta$. However, inference requires quantifying uncertainty, naturally achieved through a Bayesian framework by sampling from the posterior distribution of $\btheta$ under an appropriate prior. The next section establishes a key connection between deep GPs and dropout-based DNNs, demonstrating that the optimization in (\ref{eq:loss}) corresponds to optimizing variational posterior distribution for a prior on $\btheta$. This crucial insight establishes the theoretical backbone of efficient posterior sampling of $\btheta$ by applying random dropout to the estimated $\widehat{\btheta}$ from (\ref{eq:loss}), known as the \emph{MC-dropout} architecture \citep{gal2016bayesian}.


\subsubsection{Monte Carlo (MC) Dropout Architecture: Inference from a Deep Neural Network using Deep Gaussian Process}
This section establishes that the objective (\ref{eq:loss}) is equivalent to finding optimal parameters of a variational approximation for the posterior distribution of $\btheta$ under a deep GP prior construction provided below. This equivalence holds without simplifying assumptions, and leads to a strategy of drawing posterior samples of $\btheta$.

Consider a sequence of features $\{{\boldsymbol \psi}_{(v,j)}^{(\beta,q,l_1)}\in\mathbb{R}^{k_{l_1}^{(\beta)}}:j=1,...,J_v;\:v=1,..,V;\:l_1=1,..,L_\beta\}$ over $L_\beta$ layers, with the features in the $l_1$th layer is of dimension $k_{l_1}^{(\beta)}$. 
Let $\boldsymbol{\Psi}^{(\beta)}_{q,l_1}=\Big[{\boldsymbol \psi}_{(1,1)}^{(\beta,q,l_1)}:\cdots:{\boldsymbol \psi}_{(1,J_1)}^{(\beta,q,l_1)}:\cdots:{\boldsymbol \psi}_{(V,1)}^{(\beta,q,l_1)}:\cdots:{\boldsymbol \psi}_{(V,J_V)}^{(\beta,q,l_1)}\Big]^T$ denote a $J\times k_{l_1}^{(\beta)}$ matrix obtained by stacking features in the $l_1$th layer. Let the $k$th column of this matrix is denoted as $\boldsymbol{\psi}^{(\beta)}_{q,l_1,k}$. Assuming conditional independence of $\boldsymbol{\psi}^{(\beta)}_{q,l_1,k}$ over $k$, we model $\beta_q(\cdot)$ using a deep GP with $L_\beta$ layers. Conditional on the features in the $(l_1-1)$th layer, the realizations at the $l_1$th layer follow
\begin{align}\label{eq:psi}
\boldsymbol{\psi}^{(\beta)}_{q,l_1,k}|\boldsymbol{\Psi}^{(\beta)}_{q,l_1-1} &\sim N(\bzero, {\mathbf{\Sigma}}^{(\beta)}_{q,l_1}),\quad k=1,\cdots,k^{(\beta)}_{l_1},\:\:l_1=1,...,L_\beta, 
\end{align}
with the coefficients $\beta_q(\bs_{v,j})$ being constructed through the $L_\beta$th layer of the features. Specifically, if $\bbeta_q$ is a $J$ dimensional vector $(\beta_q(\bs_{v,j}):j=1,..,J_v;\:v=1,..,V)^T$, then we model
$\bbeta_q=\sigma_{q,L_\beta}^{(\beta)}(\boldsymbol{\Psi}^{(\beta)}_{q,L_\beta})$.
The covariance matrix ${\mathbf{\Sigma}}^{(\beta)}_{q,l_1} \in \mathbb{R}^{ J \times J}$ depends on features from the previous layer. 
Let $\boldsymbol\Phi^{(\beta)}_{q,l_1}=\Big[{\boldsymbol \phi}_{(1,1)}^{(\beta,q,l_1)}:\cdots:{\boldsymbol \phi}_{(1,J_1)}^{(\beta,q,l_1)}:\cdots:{\boldsymbol \phi}_{(V,1)}^{(\beta,q,l_1)}:\cdots:{\boldsymbol \phi}_{(V,J_V)}^{(\beta,q,l_1)}\Big]^T$ be a $J\times k_{l_1}^{(\beta)}$ dimensional matrix of feature vectors after applying transformation through the activation function, such that $\boldsymbol\phi^{(\beta,q,l_1)}_{(v,j)}=\sigma_{q,l_1}^{(\beta)}(\boldsymbol\psi^{(\beta,q,l_1)}_{(v,j)})$. The covariance matrix ${\mathbf{\Sigma}}^{(\beta)}_{q,l_1}$ is expressed as 
\begin{align}\label{eq:approx_cov}
{\mathbf{\Sigma}}^{(\beta)}_{q,l_1}=\frac{1}{k^{(\beta)}_{l_1}}\sigma^{(\beta)}_{q,l_1}(\boldsymbol\Phi^{(\beta)}_{q,l_1-1}\mathbf{W}^{(\beta)\top}_{q,l_1}+ \mathbf{b}^{(\beta)}_{q,l_1})
   \sigma^{(\beta)}_{q,l_1}(\boldsymbol\Phi^{(\beta)}_{q,l_1-1}\mathbf{W}^{(\beta)\top}_{q,l_1}+ \mathbf{b}^{(\beta)}_{q,l_1})^{\top}.
   \end{align}

\begin{figure}[htbp]
\begin{center}
\includegraphics[width = 0.8\textwidth,trim={1.5mm 1.5mm 1.5mm 1.5mm}]{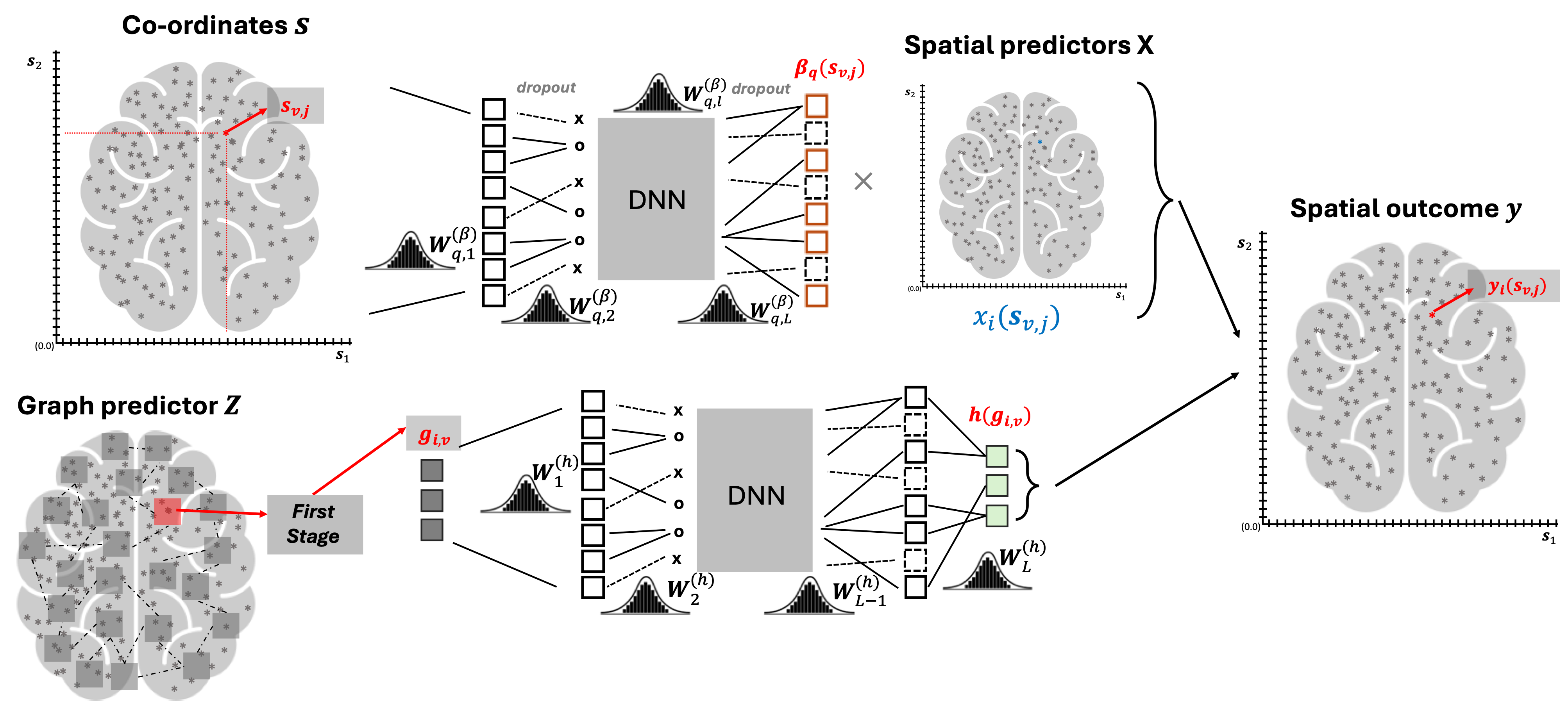}
\end{center}
\caption[]{Illustration for the XAI.}
\label{xai_illu}
\end{figure}

Similarly, define a sequence of features $\{\boldsymbol\psi^{(h,l_2)}_{(i,v)}\in \mathbb{R}^{k_{l_2}^{(h)}}: i=1,..,n;\:v=1,..,V;\:l_2=1,...,L_h\}$
and construct non-linear transformation $\{\boldsymbol{\phi}^{(h,l_2)}_{(i,v)}: i=1,..,n;\:v=1,..,V;\:l_2=1,...,L_h\}$ of these features after applying the activation function,
such that $\boldsymbol{\phi}^{(h,l_2)}_{(i,v)}=\sigma^{(h)}_{l_2}(\boldsymbol\psi^{(h,l_2)}_{(i,v)}) \in \mathbb{R}^{k^{(h)}_{l_2}}$. Collapsing the three-way tensor $\{\boldsymbol{\psi}^{(h,l_2)}_{(i,v)}\}_{i=1,v=1}^{n,V}\in\mathbb{R}^{n\times V\times k_{l_2}^{(h)}}$ along the third mode leads to a $(nV)\times k_{l_2}^{(h)}$ matrix $\boldsymbol{\Psi}^{(h)}_{l_2}$, whose $k$th column is denoted as $\boldsymbol{\psi}^{(h)}_{l_2,k}$. Assuming independence of $\boldsymbol{\psi}^{(h)}_{l_2,k}$ across $k$, we model $h(\cdot)$ using a deep GP with $L_h$ layers. Specifically, conditional on the parameters in $(l_2-1)$th layer, the finite-dimensional realizations of the $l_2$th layer follow
\begin{align}\label{eq:psih}
\boldsymbol{\psi}^{(h)}_{l_2,k}|\boldsymbol{\Psi}^{(h)}_{l_2-1} &\sim N(0, {\mathbf{\Sigma}}^{(h)}_{l_2}),\quad k=1,\cdots,k^{(h)}_{l_2},\:l_2=1,..,L_h,
\end{align}
with the non-linear function $h(\bg_{i,v})$ being constructed through the $L_h$th layer of the features, i.e., if $\bh$ is an $nV$ dimensional vector obtained by vectorizing the $n\times V$ dimensional matrix $(h(\bg_{i,v}))_{i,v=1}^{n,V}$, then we model
$\bh=\sigma_{L_h}^{(h)}(\boldsymbol{\Psi}^{(h)}_{L_h})$.
Let $\boldsymbol\Phi^{(h)}_{l_2}$ is a $(nV)\times k_{l_2}^{(h)}$ matrix formed by collapsing the $n\times V\times k_{l_2}^{(h)}$ tensor $\lbrace \boldsymbol\phi^{(h,l_2)}_{(i,v)} \rbrace_{i,v=1}^{n,V}$ in the third mode. We covariance matrix ${\mathbf{\Sigma}}^{(h)}_{l_2}$ is constructed as 
\begin{align}\label{eq:approx_cov_h}
{\mathbf{\Sigma}}^{(h)}_{l_2}=\frac{1}{k^{(h)}_{l_2}}\sigma^{(h)}_{l_2}(\boldsymbol\Phi^{(h)}_{l_2-1}\mathbf{W}^{(h)\top}_{l_2}+ \mathbf{b}^{(\beta)}_{l_2})
   \sigma^{(h)}_{l_2}(\boldsymbol\Phi^{(h)}_{l_2-1}\mathbf{W}^{(h)\top}_{l_2}+ \mathbf{b}^{(h)}_{l_2})^{\top}.   
\end{align}
Assuming that the final layer in both generative models consists of one component, i.e., $k_{L_\beta}^{(\beta)}=k_{L_h}^{(h)}=1$, 
the proposed generative model is given by,
\begin{align}\label{eq:fs_last}
& y_i(\bs_{v,j})|\{\mathbf{\Psi}^{(\beta)}_{q,L_\beta}\}_{q=1}^Q, \mathbf{\Psi}^{(h)}_{L_h},\tau^2, \bx(\bs_i),\bg_{i,v} \sim N(\bx_i(\bs_{v,j})^T\bbeta(\bs_{v,j})+h(\bg_{i,v}),\tau^2),\nonumber\\
& \beta_q(\bs_{v,j})=\sigma_{q,L_\beta}^{(\beta)}(\psi_{(v,j)}^{(\beta,q,L_\beta)}),\:h(\bg_{i,v})=\sigma_{L_h}^{(h)}(\psi_{(i,v)}^{(h,L_h)}),\:i=1,..,n;\:q=1,..,Q;\:j=1,..,J_v;\:v=1,...,V\nonumber\\
&\qquad \boldsymbol{\psi}^{(\beta)}_{q,l_1,k}|\boldsymbol{\Psi}^{(\beta)}_{q,l_1-1} \sim N(\bzero, {\mathbf{\Sigma}}^{(\beta)}_{q,l_1}),\:q=1,..,Q;\:k=1,...,k_{l_1}^{(\beta)};\:l_1=1,\dots,L_\beta\nonumber\\
&\qquad\qquad \boldsymbol{\psi}^{(h)}_{l_2,k}|\boldsymbol{\Psi}^{(h)}_{l_2-1} \sim N({\boldsymbol 0}, {\mathbf{\Sigma}}^{(h)}_{l_2}),\quad k=1,\cdots,k^{(h)}_{l_2},\:l_2=1,..,L_h,.
\end{align} 
Figure~\ref{xai_illu} illustrates the proposed explainable AI (XAI) approach.
We adopt variational Bayesian inference called \emph{MC dropout}, to approximate posterior distribution of $\btheta$.
Let $\mathbf{D}= \lbrace\lbrace\lbrace \mathbf{x}_{i}(\bs_{v,j}),\mathbf{s}_{v,j},\mathbf{g}_{i,v} \rbrace_{j=1}^{J_v}\rbrace_{v=1}^{V}\rbrace_{i=1}^{n}$ denote the input data. For Bayesian inference, independent standard normal priors are set on each component of the weight and bias parameters, denoted by $p(\btheta)=N({\boldsymbol 0},\bI)$. 
We construct a variational approximation of the posterior distribution of parameters $\tilde{q}(\boldsymbol\theta)$, constructed as $\tilde{q}(\boldsymbol{\theta}):=\prod_{l_1=1}^{L_\beta}\prod_{q=1}^Q\tilde{q}(\mathbf{W}^{(\beta)}_{q,l_1},\mathbf{b}^{(\beta)}_{q,l_1})\prod_{l_2=1}^{L_h}\tilde{q}(\mathbf{W}^{(h)}_{l_2},\mathbf{b}^{(h)}_{l_2})$ where each term represents the variational distribution of the corresponding weight matrices and bias vectors. Let  $\bw_{q,l_1,t}^{(\beta)}=(w_{l,j,tp}^{(\beta)}:p=1,...,k_{l_1-1}^{(\beta)})$ and $\bw_{l_2,t}^{(h)}=(w_{l_2,tp}^{(h)}:p=1,...,k_{l_2-1}^{(h)})$ be the $t$th row of the matrices  $\mathbf{W}^{(\beta)}_{q,l_1}$ and $\mathbf{W}^{(h)}_{l_2}$, respectively.  The variational distribution is constructed as 
\begin{align}\label{variationaldist}
   & \tilde{q}(\mathbf{W}^{(\beta)}_{q,l_1},\mathbf{b}^{(\beta)}_{q,l_1})  = \prod_{\forall t} \tilde{q}(\bw^{(\beta)}_{q,l_1,t},b^{(\beta)}_{q,l_1,t}),\:\: \tilde{q}(\mathbf{W}^{(h)}_{l_2},\mathbf{b}^{(h)}_{l_2})  = \prod_{\forall t} \tilde{q}(\bw^{(h)}_{l_2,t},b^{(h)}_{l_2,t})\nonumber\\
   &\tilde{q}(\bw^{(\beta)}_{q,l_1,t},b^{(\beta)}_{q,l_1,t}) = p^{(\beta)}_{l_1}N((\bmu^{w,(\beta)}_{q,l_1,t},\mu^{b,(\beta)}_{q,l_1,t})^\top,\sigma^2\bI)+(1-p^{(\beta)}_{l_1})N(\mathbf 0,\sigma^2\bI),\nonumber\\
  &\tilde{q}(\bw^{(h)}_{l_2,t},b^{(h)}_{l_2,t}) = p^{(h)}_{l_2}N((\bmu^{w,(h)}_{l_2,t},\mu^{b,(h)}_{l_2,t})^\top,\sigma^2\bI)+(1-p^{(h)}_{l_2})N(\mathbf 0,\sigma^2\bI),
\end{align}
where $l_1=1,...,L_\beta$ and $l_2=1,...,L_h$. The parameters $w^{(\beta)}_{q,l_1,tp}$ and $w^{(h)}_{l_2,tp}$ are the $(t,p)$th element of the weight matrices $\mathbf{W}^{(\beta)}_{q,l_1}$ and $\mathbf{W}^{(h)}_{l_2}$, respectively. Similarly, $b^{(\beta)}_{q,l_1,t}$ and $b^{(h)}_{l_2,t}$ are the $t$th element of the bias vectors $\mathbf{b}^{(\beta)}_{q,l_1}$ and $\mathbf{b}^{(h)}_{l_2}$, respectively. Each scalar weight or bias has a variational distribution of the Gaussian mixture, with each mixing component having a small variance $\sigma^2$. Let 
$\bDelta$ denote the collection of variational parameters $\{\bmu_{q,l_1,t}^{w,(\beta)}= (\mu_{q,l_1,tp}^{w,(\beta)} :\:p=1,..,k_{l_1-1}^{(\beta)}): l_1=1,..,L_\beta; \:t=1,..,k_{l_1}^{(\beta)};\:q=1,..,Q\} $, $\{\bmu_{l_2,t}^{w,(h)}=(\mu_{l_2,tp}^{w,(h)}:\:p=1,..,k_{l_2-1}^{(h)}):  l_2=1,..,L_h;\:t=1,..,k_{l_2}^{(h)}\}$, $\{\mu_{q,l_1,t}^{b,(\beta)}:\:l_1=1,..,L_\beta ;\:t=1,..,k_{l_1}^{(\beta)};\:q=1,..,Q\}$, $\{\mu_{l_2,t}^{b,(h)}:l_2=1,..,L_h;\:t=1,..,k_{l_2}^{(h)}\}$ and $\sigma^2$.
Under the mixture normal specification for the variational distribution of all weight and bias parameters, as the inclusion probabilities $p^{(\beta)}_{l_1},p^{(h)}_{l_2} \in [0,1]$ approaches 0, $\tilde{q}({\bw}^{(\beta)}_{q,l_1,t})$ and $\tilde{q}({\bw}^{(h)}_{l_2,t})$ becomes close to $N(\mathbf 0,\sigma^2\bI)$. We denote the variational distribution as $\tilde{q}(\btheta|\bDelta)$ to explicitly show its dependence on the variational parameters. The optimal variational parameters are set by maximizing 
\begin{align}\label{var_elbo}
\mbox{E}_{\tilde{q}}[{\log(\pi(\boldsymbol{\theta},\mathbf{D}))}]-\mbox{E}_{\tilde{q}}[\log \tilde{q}({\boldsymbol{\theta}}|\bDelta)],
\end{align}
the evidence lower bound (ELBO), where $\pi(\boldsymbol{\theta},\mathbf{D})$ is the joint distribution of data $\bD$ and parameters $\btheta$. 
Let $\btheta^{(m)}=\{\lbrace \mathbf{W}_{q,l_1}^{(\beta)(m)}, \mathbf{b}_{q,l_1}^{(\beta)(m)}\rbrace_{l_1=1,q=1}^{L_\beta, Q}$, $\lbrace \mathbf{W}_{l_2}^{(h)(m)}, \mathbf{b}_{l_2}^{(h)(m)} \rbrace_{l_2=1}^{L_h}\}$ be MC samples from the variational distribution in \eqref{variationaldist}. 
The detailed derivations in Section 1 of the supplementary file show that an MC approximation of ELBO is given by,
\begin{align}\label{GPMCKLsuppl}
&\mathcal{L}_{\text{GP-MC}}(\tau^2,\bDelta) \approx \frac{1}{M}\sum_{m=1}^{M}\sum_{i=1}^{n} \sum_{v=1}^{V}\sum_{j=1}^{J_v} \log p(\mathbf{y}_{i}(\bs_{v,j})|\bx_{i}(\bs_{v,j}),\bg_{i,v},
\tau^2,\btheta^{(m)})- \nonumber \\
&\sum_{l_{1}=1}^{L_\beta}\sum_{q=1}^{Q}\frac{p^{(\beta)}_{l_1}}{2}(||\bmu_{q,l_{1}}^{w,(\beta)}||_2^2)-\sum_{l_{1}=1}^{L_\beta}\sum_{q=1}^{Q}\frac{p^{(\beta)}_{l_1}}{2}(||\bmu_{q,l_{1}}^{b,(\beta)}||_2^2)-\sum_{l_{2}=1}^{L_h}\frac{p^{(h)}_{l_2}}{2}(||\bmu_{l_{2}}^{w,(h)}||_2^2)-\sum_{l_{2}=1}^{L_h}\frac{p^{(h)}_{l_2}}{2}(||\bmu_{l_{2}}^{b,(h)}||_2^2).
\end{align}
The data likelihood $p(\mathbf{y}_{i}(\bs_{v,j})|\bx_{i}(\bs_{v,j}),\bg_{i,v},
\tau^2,\btheta^{(m)})$ is given by the normal distribution in the first row of \eqref{eq:fs_last}.\\
\underline{\textbf{Approximating posterior samples of DNNs using MC dropout.}}
By setting $\lambda_{l_1}^{(W, \beta_q)},$\\ $\lambda_{l_1}^{(b, \beta_q)}, \lambda_{l_2}^{(W, h)}$, and $\lambda_{l_2}^{(b, h)}$ as $\frac{p^{(\beta)}_{l_1}}{2N}$, $\frac{p^{(\beta)}_{l_1}}{2N}$, $\frac{p^{(h)}_{l_2}}{2N}$ and $\frac{p^{(h)}_{l_2}}{2N}$, respectively, the negative of the objective function \eqref{GPMCKLsuppl} after scaling with $1/N$ becomes equivalent to \eqref{eq:loss}. Details of the derivation are provided in Section 1 of the supplementary file. 
This reveals that finding optimal $\widehat{\btheta}$ from a DNN with dropout by minimizing the objective function (\ref{eq:loss}), is mathematically equivalent to estimating the optimal variational parameters $\widehat{\bDelta}$ by maximizing the objective function \eqref{GPMCKLsuppl}. This is a
highly consequential result, as it allows efficient estimation of variational parameters $\widehat{\bDelta}$ by optimizing the loss function in (\ref{eq:loss}) using an efficient stochastic gradient descent (SGD) algorithm. Assuming $\sigma^2\approx 0$, approximate posterior samples of the weights and biases $\btheta$ can then be drawn by applying dropout to $\widehat{\btheta}$. To elaborate on it, for each layer $l_1$, generate vectors of Bernoulli random variables $\bz^{(\beta)}_{q,l_1}\in\{0,1\}^{k^{(\beta)}_{l_1}}$ with keep probability $1-p^{(\beta)}_{l_1}$. Then construct a dropout mask of binary elements $\mathbf{Z}_{q,l_1}^{(\beta)}=\bz^{(\beta)}_{q,l_1}\mathbf 1^{\top}_{k^{(\beta)}_{l-1}}\in\mathbb{R}^{k_{l_1}^{(\beta)}\times k_{l_1-1}^{(\beta)}}$ for the $l_1$th layer for weights corresponding to DNN architecture of $\beta_q(\cdot)$.  Similarly, each layer $l_2$, generate vectors of  Bernoulli random variables $\bz^{(h)}_{q,l_2}\in\{0,1\}^{k^{(h)}_{l_2}}$, and construct each element of the dropout mask $\mathbf{Z}_{l_2}^{(h)}=\bz^{(h)}_{q,l_2}\mathbf 1^{\top}_{k^{(h)}_{l-1}}\in\mathbb{R}^{k_{l_2}^{(h)}\times k_{l_2-1}^{(h)}}$ for $l_2$th layer for weights for the DNN construction of $h(\cdot)$. We apply dropout masks $\mathbf{Z}_{q,l_1}^{(\beta)}$, $\mathbf{z}_{q,l_1}^{(\beta)}$, $\mathbf{Z}_{l_2}^{(h)}$ and $\mathbf{z}_{l_2}^{(h)}$ to the estimated weight matrices and bias vectors $\widehat{\bW}_{q,l_1}^{(\beta)}$, $\widehat{\bb}_{q,l_1}^{(\beta)}$, $\widehat{\bW}_{l_2}^{(h)}$ and $\widehat{\bb}_{l_2}^{(h)}$ (obtained by optimizing \eqref{eq:loss}), respectively, to obtain $\widehat{\bW}_{q,l_1}^{(\beta)}\odot \mathbf{Z}_{q,l_1}^{(\beta)}$, $\widehat{\bb}_{q,l_1}^{(\beta)}\odot \mathbf{z}_{q,l_1}^{(\beta)}$, $\widehat{\bW}_{l_2}^{(h)}\odot \mathbf{Z}_{l_2}^{(h)}$ and $\widehat{\bb}_{l_2}^{(h)}\odot \mathbf{z}_{l_2}^{(h)}$. 
Given $\sigma^2\approx 0$, these corresponds to approximate draws from the marginal variational posteriors of  $\bW_{q,l_1}^{(\beta)}$, $\bb_{q,l_1}^{(\beta)}$, $\bW_{l_2}^{(h)}$ and $\bb_{l_2}^{(h)}$ in (\ref{variationaldist}), respectively.

Repeating this procedure $F$ times yields $F$ approximate posterior samples from the DNNs, which are further used for posterior inference of model parameters and posterior predictive inference.
Details of the algorithm are presented in Algorithm 1.

\begin{algorithm}[h]
\caption{Explainable AI Algorithm}
\label{alg:xai}

\textbf{Step 1:} Initialize parameters $\btheta$ at $\btheta^{(0)}$\;

\vspace{0.2em}
\textbf{Step 2:} Find the optimal point estimates $\btheta$ using SGD\;
\For{$\text{iter} = 1$ \KwTo $n_{\text{iter}}$}{
    Update $\btheta$ using SGD to minimize loss function in Eq.~\eqref{eq:loss}\;
}
\textbf{Return:} Optimized estimates 
$\widehat{\btheta} = \left\{ 
\begin{array}{l}
\{\widehat{\bW}^{(\beta)}_{q,l_1}, \widehat{\bb}^{(\beta)}_{q,l_1}\}_{l_1=1,q=1}^{L_\beta, Q}, \\
\{\widehat{\bW}^{(h)}_{l_2}, \widehat{\bb}^{(h)}_{l_2}\}_{l_2=1}^{L_h}
\end{array}
\right\}$, $\widehat{\beta}_0$\;

\vspace{0.2em}
\textbf{Step 3:} Draw samples of $\btheta$ and $\tau^2$ using dropout\;
\For{$f = 1$ \KwTo $F$}{
    Sample dropout masks:\;
  For $l_1=1,..,L_\beta$, $q=1,..,Q$, sample node-wise dropout mask $\mathbf{z}^{(\beta)(f)}_{q,l_1}\in\mathbb{R}^{k^{(\beta)}_{l_1}}$ from $\text{Bernoulli}(1 - p^{(\beta)}_{l_1})$ and set
    $\mathbf{Z}_{q,l_1}^{(\beta)(f)}=\mathbf{z}^{(\beta)(f)}_{q,l_1}{\boldsymbol 1}_{k_{l_1-1}^{(\beta)}}^\top$.\\
  Similarly, for $l_2=1,..,L_h$, sample node-wise dropout mask $\mathbf{z}^{(h)(f)}_{l_2}\in\mathbb{R}^{k_{l_2}^{(h)}}$ from $ \text{Bernoulli}(1 - p^{(h)}_{l_2})$ and set $\mathbf{Z}_{l_2}^{(h)(f)}=\mathbf{z}^{(h)(f)}_{l_2}{\boldsymbol 1}_{k_{l_1-1}^{(h)}}^\top$.
    
    Apply masks element-wise to $\widehat{\btheta}$ to obtain sparse $\btheta^{(f)}$\; i.e., set $\widehat{\bW}_{q,l_1}^{(\beta)}\odot \mathbf{Z}_{q,l_1}^{(\beta)(f)}$, $\widehat{\bb}_{q,l_1}^{(\beta)}\odot \mathbf{z}_{q,l_1}^{(\beta)(f)}$, $\widehat{\bW}_{l_2}^{(h)}\odot \mathbf{Z}_{l_2}^{(h)(f)}$ and $\widehat{\bb}_{l_2}^{(h)}\odot \mathbf{z}_{l_2}^{(h)(f)}$.

    Compute $\beta_q(\bs_{v,j})^{(f)}$ and $h(\bg_{i,v})^{(f)}$ using Eq.~\eqref{eq:XAI} from $\btheta^{(f)}$\;

    Compute variance:
    $\tau^{2(f)} = \frac{1}{n\sum_v J_v} \sum_{i=1}^n\sum_{v=1}^V\sum_{j=1}^{J_v} \left( y_i(\bs_{v,j}) - \widehat{\beta}_0 - \bx_i(\bs_{v,j})^T \bbeta^{(f)}(\bs_{v,j}) - h^{(f)}(\bg_{i,v}) \right)^2$\;
}

\vspace{0.2em}
\textbf{Step 4:} \Return{Approximate posterior samples: 
$\{\btheta^{(1)}, \dots, \btheta^{(F)}\}$, 
$\{\tau^{2(1)}, \dots, \tau^{2(F)}\}$\;}

\end{algorithm}

\subsection{Explainable Inference with Deep Neural Network}
A key feature of XAI is ability to quantify uncertainty in the estimation of the spatially-varying image predictor coefficients $\beta_q(\cdot)$, the function representing node effects $h(\cdot)$, and in the prediction of the outcome image, thereby enabling \emph{explainable inference} within deep neural networks. Let $\bD=\{(y_i(\bs_{v,j}),\bx_i(\bs_{v,j}),\bg_{i,v})^T:i=1,..,n;\:j=1,..,J_v;\:v=1,..,V\}$ denote the observed data. To predict unobserved outcome $y_{i*}(\bs_{v,j})$ for a new sample indexed $i^*$, we first employ the algorithm outlined in Section 3.2.1 to construct the features $\{\bg_{i^*,v}:\:v=1,..,V\}$ from the graph predictor $\bZ_{i^*}$ for the new sample.  The posterior predictive distribution is then given by
$p(y_{i^*}(\bs_{v,j})|\bD)=\int\cdots\int  N(y_{i^*}(\bs_{v,j})|\bx_{i^*}(\bs_{v,j})^T\bbeta(\bs_{v,j})+h(\bg_{i^*,v}),\tau^2)\pi(\btheta|\bD)d\btheta\:d\tau^2.$
Since it is challenging to compute the posterior predictive distribution directly, we approximate it through the samples of $\btheta,\tau^2$. 
With approximate posterior samples $\lbrace\btheta^{(f)},\tau^{2(f)}\rbrace_{f=1}^F$ drawn following Algorithm 1, we adopt composition sampling, wherein the $f$th post burn-in iterate 
$y_{i^*}^{(f)}(\bs_{v,j})$ is drawn from $N(y_{i^*}(\bs_{v,j})|\bx_{i^*}(\bs_{v,j})^T\bbeta(\bs_{v,j})^{(f)}+h(\bg_{i^*,v})^{(f)},\tau^{2(f)})$. 

Inference on $h(\bg_{i,v})$ is based on approximate posterior samples of $\{h(\bg_{i,v})^{(f)}\}_{f=1}^F$ constructed from \eqref{eq:XAI} using samples $\{\btheta^{(f)}\}_{f=1}^F$.
Although $\beta_q(\bs_{v,j})$ can be directly inferred from its approximate posterior samples $\{\beta_q(\bs_{v,j})^{(f)}:f=1,..,F\}$, we observe slight under-coverage in the resulting credible intervals. Thus, we approximate its posterior with a normal distribution, using empirical mean $\mu_q(\bs_{v,j})= \frac{1}{F}\sum_{f=1}^{F}\beta_q(\bs_{v,j})^{(f)}$ and variance
 $\sigma^2_q(\bs_{v,j}) = \frac{1}{F-1}\sum_{f=1}^{F}(\beta_q(\bs_{v,j})^{(f)}- \mu_q(\bs_{v,j}))^2$,  and generate refined samples from $N(\mu_q(\bs_{v,j}), \sigma^2_q(\bs_{v,j}))$.


\section{Simulation Study}
This section evaluates the performance of XAI alongside relevant Bayesian linear, non-linear, and deep learning methods for image-on-image regression, focusing on three key aspects: predictive inference, inference on the spatially-varying predictor coefficients $\bbeta_q(\cdot)$, and inference on $h(\cdot)$, which captures the non-linear effect of the network predictor. 

\subsection{Simulation Design}
In all simulations, the spatial locations $\bs_{v,j}$ are assumed to be three-dimensional ($d=3$) and are drawn uniformly from the domain $[0,2]\times [0,2]\times [0,2]$. Each simulation assumes $Q=2$, sets $\bx_i(\bs_{v,j})=\bx_i\in\mathbb{R}^Q$ for all $v,j$, and draws $\bx_i$ from $N(0,\bI_2)$. 
The spatial predictor coefficients in $\beta_q(\cdot)$ ($q=1,2$) are simulated from a Gaussian process (GP) with mean zero and an exponential covariance function, characterized by the spatial variance $\delta_{\beta,q}^2$ and the spatial scale parameter $\eta_{\beta,q}$. Specifically, Cov($\beta_q(\bs_{v,j}),\beta_q(\bs_{v',j'}))=\delta_{\beta,q}^2\exp(-||\bs_{v,j}-\bs_{v',j'}||_2/\eta_{\beta,q})$, for any $v,v'=1,...,V$, $j=1,...,J_v$ and $j'=1,...,J_{v'}$. The latent network effects $\bg_{i,v}$'s are simulated from $N(0,\bI_2)$. The regression function $h(\cdot)$ for the latent network effect is modeled as a Gaussian process with mean zero and an exponential covariance kernel, defined by the spatial variance parameter $\delta_h^2$ and spatial scale parameter $\eta_h$.
Each simulation uses a sample size of $300$ with low signal-to-noise ratio (SNR) to reflect real neuroimaging data. The response is generated using equation (\ref{additive_reg}) with noise variance $\tau^2 = 4$. Of the $300$ samples, $180$ are used for training, $60$ for validation, and $n^*=60$ for testing. 
Simulations assume equal regions per node, i.e., $J_1=\cdots=J_V$, and fix the number of network nodes at $V = 30$.

We explore simulation scenarios by varying the total number of regions $J$. Additionally, different combinations of $(\delta_{\beta,1}^2,\delta_{\beta,2}^2,\eta_{\beta,1},\eta_{\beta,2},\delta_h^2,\eta_h) $ induce varying SNR levels, allowing us to assess model performance under a range of SNRs. Simulation results on the effect of varying correlation parameters are presented in Section~2 of the supplementary file.

\subsection{Competitors and Metric of Comparison} 
We implement the proposed XAI approach in {\tt TensorFlow} with ReLU activation function in all hidden layers and run all experiments on a 4-core, 2.3 GHz Intel Core i7 processor. The objective function in (\ref{eq:loss}) is optimized using the \texttt{Adam} optimizer with an initial learning rate of $10^{-2}$, a decay rate of $0.97$, and decay steps of 10,000. For the scenario that varies the total number of ROIs, we use a batch size of 10 and a maximum of 300 training epochs, whereas for the scenario that varies the signal–to–noise ratios, the batch size is 20 with a maximum of 500 epochs. We apply early stopping based on the validation mean squared error, using a patience of 8 epochs and a minimum improvement threshold of $10^{-4}$, and we restore the weights corresponding to the best validation performance. To evaluate performance, we compare XAI's predictions and estimation of $\beta_q(\cdot)$ and $h(\cdot)$ against Bayesian Additive Regression Trees (BART) \citep{chipman2010bart}, BIRD-GP \citep{ma2023bayesian}, and a Generalized Linear Model (GLM). For fair comparison, BART includes spatial coordinates along with imaging and network inputs. While GLM is commonly used in image-on-image regression \citep{batouli2020some}, it lacks non-linear modeling and uncertainty quantification. BART also ignores structural differences by stacking network and spatial objects into a single input. For BIRD–GP, we use 30 basis functions and optimize the model with 2,500 gradient steps, while for BART, we employ an ensemble of 200 trees across all scenarios.

To evaluate predictive performance, we report root mean square prediction error (RMSPE), empirical coverage, and average length of 95\% prediction intervals over $n^*$ test samples. Inference on $\beta_q(\cdot)$ and $h(\cdot)$ is assessed with root mean square error (RMSE), coverage, and length of 95\% credible intervals. Inference on $\beta_q(\cdot)$ and $h(\cdot)$ is available only for XAI and GLM, as BART and BIRD-GP do not support it. Each simulation is repeated 100 times.

\subsection{Simulation Results}
\subsubsection{\texorpdfstring{Impact of Changing the Total Number of ROIs ($J$)}{Impact of Changing the Total Number of ROIs (J)}}\label{subsec:5.3.1}

To evaluate the impact of the number of ROIs, we consider four simulation cases by varying the total number of ROIs $J$ between 150 to 1200 across four cases (see Table~\ref{tab:changeJ}). For all four cases, the variance parameters and scale parameters are kept fixed at $(\delta^2_{\beta,1}, \delta^2_{\beta,2}, \delta^2_h)=(2,1,2.5)$ and $(\eta_{\beta,1}, \eta_{\beta,2}, \eta_h)=(5,4,6)$. 

\begin{figure}[ht]
\centering\includegraphics[scale=0.65]{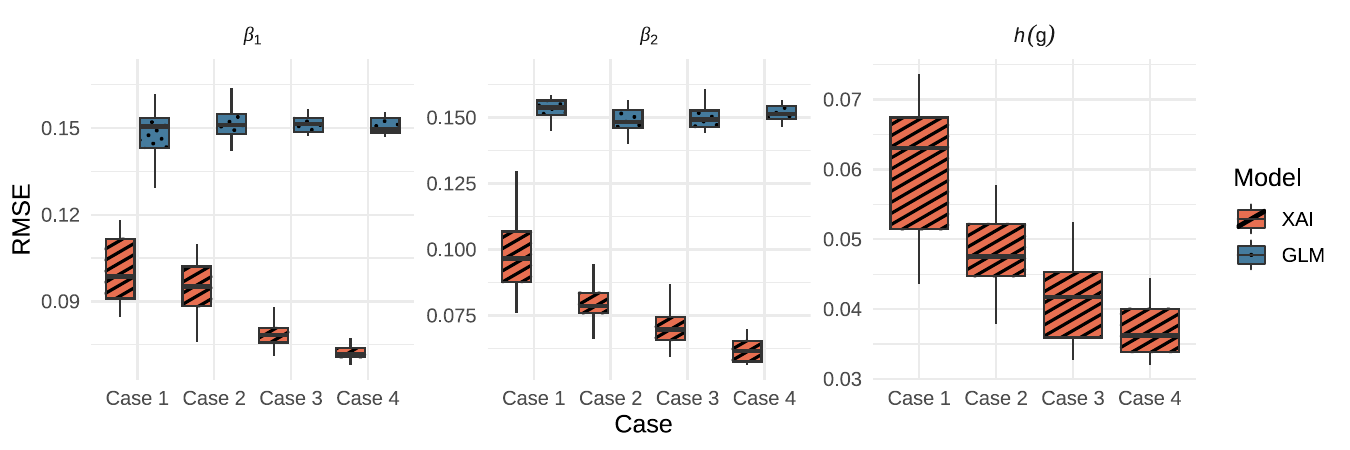}
    \caption{Root mean squared error (RMSE) comparison between the proposed XAI method and GLM for estimating the spatially varying coefficients $\beta_1(\bs)$ and $\beta_2(\bs)$, and the network latent effect $h(\bg)$ under Case 1 to Case 4. Boxplots show the distribution of RMSE across $100$ simulation replications. The inference on $\beta_1(\bs),\beta_2(\bs),h(\bg)$ are not available for BART and BIRD-GP. The inference on $h(\bg)$ is not available for GLM.}
\label{fig:CaseA_betas}
\end{figure}
Figure~\ref{fig:CaseA_betas} shows that XAI consistently yields lower RMSE than GLM across varying values of $J$ when estimating $\beta_1(\bs)$, $\beta_2(\bs)$ and $h(\bg)$. While GLM assumes a linear relationship between the network latent effect and the outcome, XAI flexibly captures complex nonlinear dependencies through its estimation of the nonlinear function $h(\bg)$. Furthermore, the GLM framework fits separate regression coefficients for each ROI without leveraging spatial correlations, resulting in stable but less adaptive point estimates with decreasing standard errors as $J$ increases. In contrast, XAI exhibits notable improvements in estimation accuracy with larger $J$, owing to its ability to borrow information spatially for estimating $\beta_1(\bs)$ and $\beta_2(\bs)$, and across the network latent space for $h(\bg)$.

\begin{figure}[h!]
\centering\includegraphics[scale=0.55]{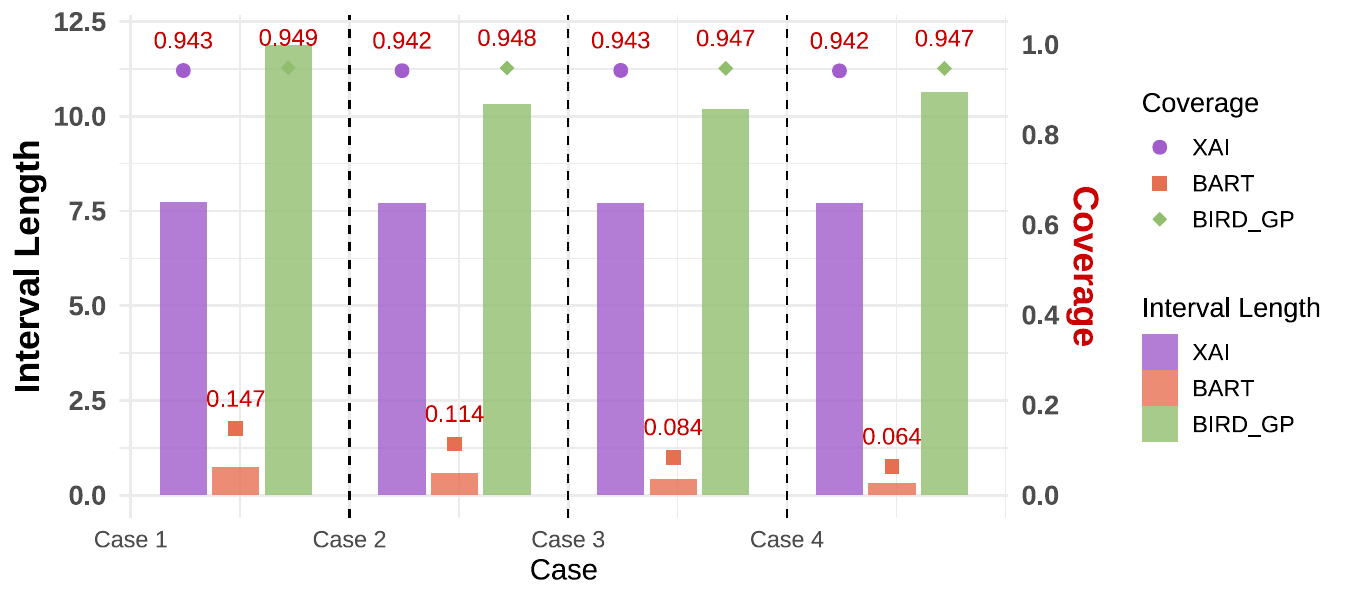}
    \caption{Coverage and length of 95\% predictive intervals averaged over replications for XAI, BIRD-GP and BART. They are not available for GLM. Results show near nominal coverage of XAI with smaller predictive intervals than BIRD-GP.}
\label{fig:CaseA}
\end{figure}

\begin{table}[h]
\centering
\caption[]{Prediction performance comparison for simulated datasets under Case 1 to Case 4. The table reports the root mean squared prediction error (RMSPE) with standard errors in parentheses for each method (XAI, BART, BIRD-GP, and GLM). The standard errors are presented over $100$ replications. Best performing model is boldfaced in each case.}\label{tab:changeJ}
\begin{tabular}{clcccc}
$J$ & \textbf{Cases}  & XAI & BART & BIRD-GP & GLM \\
\hline\hline
 150  & Case 1 & \textbf{2.0101(0.011)} &  2.0148(0.010) & 3.0234(1.0993) & 2.0264(0.011) \\
300   & Case 2 & \textbf{2.0020(0.011)}  & 2.0093(0.010) & 2.6435(0.4559) & 2.0247(0.007) \\
 600  & Case 3 & \textbf{2.0024(0.008)}  & 2.0086(0.005) & 2.6155(0.4028) & 2.0267(0.005) \\
1200   & Case 4 & \textbf{2.0038(0.007)} & 2.0065(0.004) & 2.7359(0.6534) & 2.0246(0.004) \\
\hline
\end{tabular}
\end{table}
Table~\ref{tab:changeJ} presents RMSPE for the competing models, averaged over 100 replications, with standard errors provided in parentheses. Across varying values of $J$, the point prediction performance of XAI, BART, and GLM remains stable. XAI achieves comparable accuracy to BART and GLM, while significantly outperforming the deep learning alternative BIRD-GP. As shown in Figure~\ref{fig:CaseA}, both XAI and BIRD-GP attain near-nominal coverage; however, XAI produces noticeably narrower predictive intervals. A closer inspection reveals that all four simulation scenarios assume a signal-to-noise ratio (SNR) below 1—reflective of the conditions in the real data application (Section~\ref{sec:data_analysis})—with a sample size of $n=300$. Due to its greater data requirements, BIRD-GP performs substantially worse than XAI under these settings, although the performance gap narrows as 
$n$ and SNR increase. Meanwhile, BART severely underestimates uncertainty, yielding the narrowest predictive intervals and the lowest coverage among all competing methods.

\subsubsection{Impact of Changing Signal-to-noise Ratios (SNRs)}\label{subsec:5.3.2}
To investigate the effect of SNRs, we design simulation scenarios by varying the variance parameters $\delta_{\beta,1}^2,\delta_{\beta,2}^2$ and $\delta_h^2$ corresponding to the regression functions $\beta_1(\cdot)$, $\beta_2(\cdot)$, and $h(\cdot)$ while fixing the total number of ROIs at $J=600$ and the number of nodes at $V=30$. As shown in Table~\ref{changeSNR}, increasing SNR corresponds to stronger spatial signals relative to noise.
\begin{figure}[h!]
\centering\includegraphics[scale=0.65]{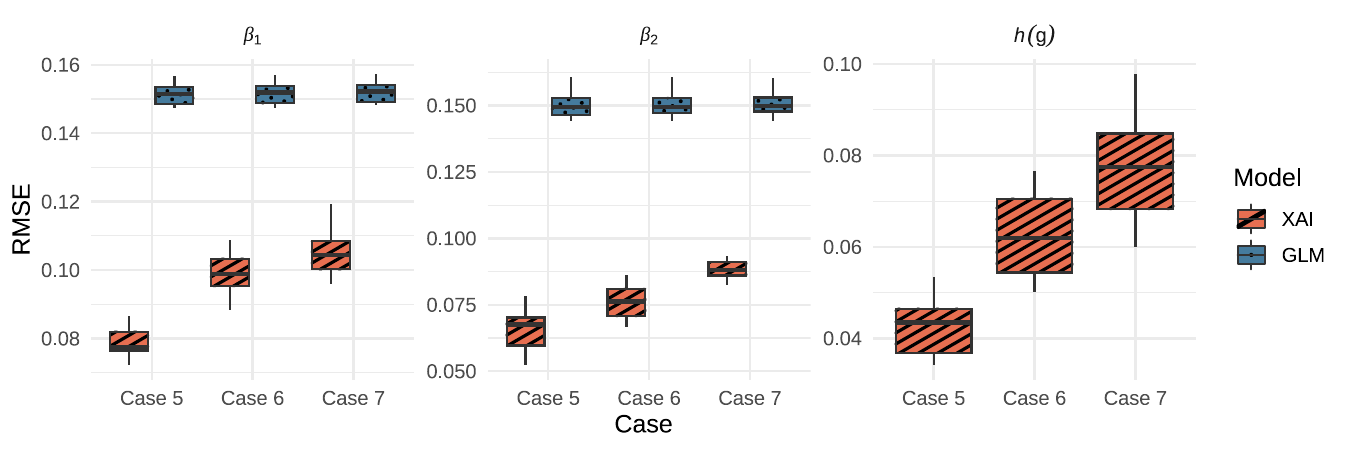}
    \caption{Root mean squared error (RMSE) comparison between the proposed XAI method and GLM for estimating the spatially varying coefficients $\beta_1(\bs)$ and $\beta_2(\bs)$, and the network latent effect $h(\bg)$ under Case 5 to Case 7. Boxplots show the distribution of RMSE across $100$ simulation replications. The inference on $\beta_1(\bs),\beta_2(\bs),h(\bg)$ are not available for BART and BIRD-GP. The inference on $h(\bg)$ is not available for GLM.}
\label{fig:CaseB_betas}
\end{figure}

\begin{figure}[h!]
\centering\includegraphics[scale=0.55]{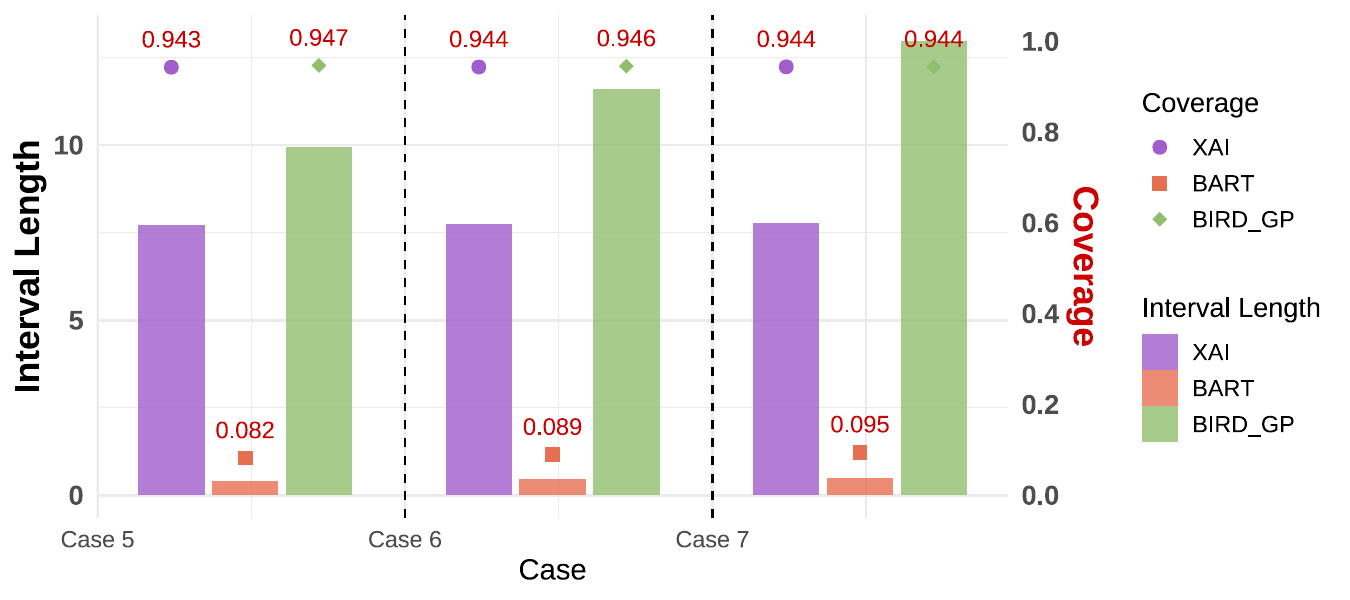}
    \caption{Coverage and length of 95\% predictive intervals averaged over replications for XAI, BIRD-GP and BART. They are not available for GLM. Results show near nominal coverage of XAI with smaller predictive intervals than BIRD-GP.}
\label{fig:CaseB}
\end{figure}

\begin{table}[h]
\centering
\caption[]{Prediction performance comparison for simulated datasets under Case 5 to Case 7 for varying signal-to-noise ratio (SNR), by varying the variance parameters $\delta_{\beta,1}^2,\delta_{\beta,2}^2,\delta_h^2$, while fixing the number of nodes ($V$) and the total number of ROIs ($J$). The table reports the root mean squared prediction error (RMSPE) with standard errors in parentheses for each method (XAI, BART, BIRD-GP and GLM). The standard errors are presented over $100$ replications. Best performing model is boldfaced in each case.}\label{changeSNR}
\begin{tabular}{cccclcccc}
$\delta_{\beta,1}^2$  & $\delta_{\beta,2}^2$ & $\delta_h^2$ & SNR &\textbf{Cases}  & XAI &  BART & BIRD-GP & GLM \\
\hline\hline
2 & 1 & 1 & 0.42  & Case 5 & \textbf{2.0020(0.009)}  & 2.0080(0.005) & 2.5549(0.382) & 2.0267(0.005) \\
4 & 3 & 3 & 0.69  & Case 6 & \textbf{2.0037(0.008)}  & 2.0120(0.005) & 2.9983(0.646) & 2.0312(0.006) \\
5 & 6 & 5 & 0.81  & Case 7 & \textbf{2.0052(0.008)} & 2.0155(0.005) & 3.3656(0.782) & 2.0358(0.008) \\
\hline\hline
\end{tabular}
\end{table}
Figure~\ref{fig:CaseB_betas} highlights the superior performance of XAI over GLM in estimating the varying coefficients $\beta_1(\bs)$, $\beta_2(\bs)$ and the non-linear function encoding network latent effect $h(\bg)$. With changing SNR, GLM’s performance remains relatively unchanged, and the performance gap between GLM and XAI tends to get smaller with increasing SNR from Case 5 to 7. Since Cases 5–7 correspond to overall low SNR settings, XAI substantially outperforms its deep learning competitor BIRD-GP in terms of point prediction, as reflected in Table~\ref{changeSNR}. Additionally, both BIRD-GP and XAI achieve near-nominal predictive coverage, although BIRD-GP produces credible intervals nearly twice as wide as those of XAI, as shown in Figure~\ref{fig:CaseB}. BART provides comparable point prediction but severely underestimates uncertainty.

\begin{figure}[h!]
\centering\includegraphics[scale=0.27]{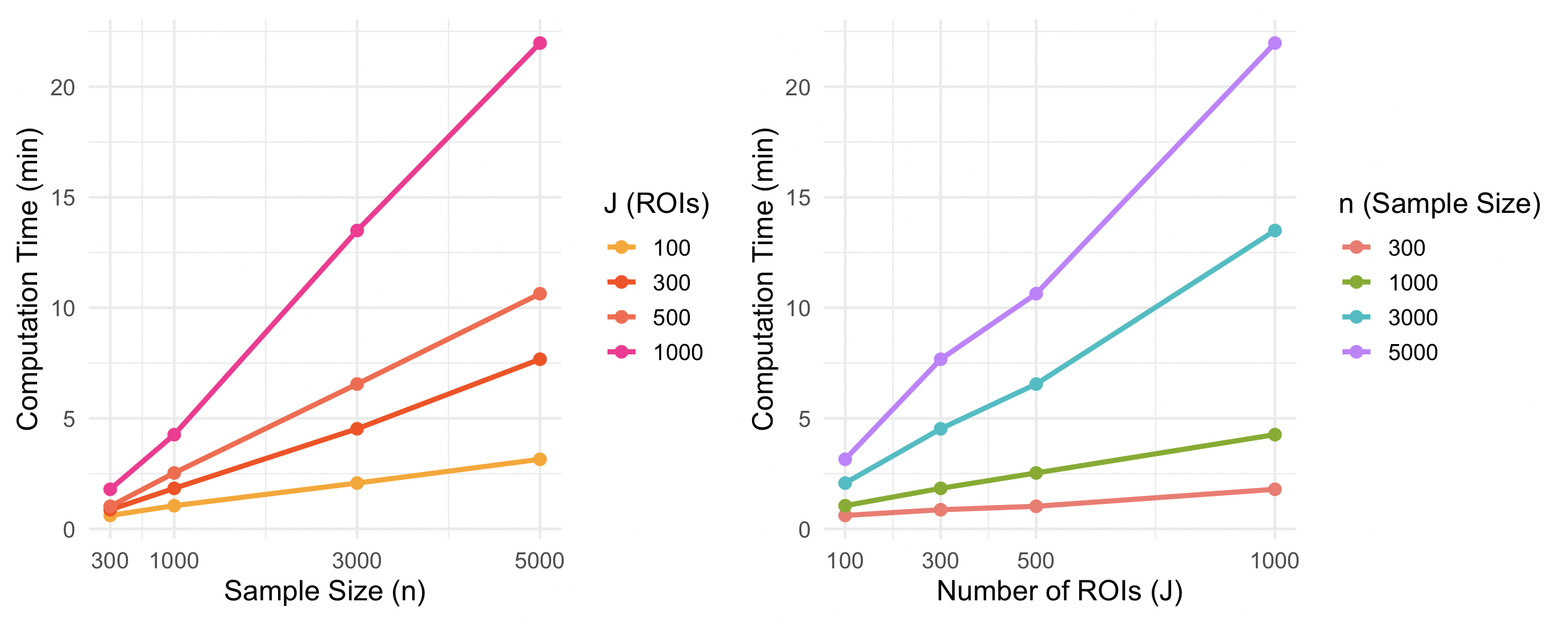}
    \caption{Computation time (in minutes) for XAI under varying dimensions of input data.
Left: Computation time as a function of sample size $n$ for different numbers of ROIs $J$. Right: Computation time as a function of the number of ROIs $J$ for different sample size $n$.}
\label{fig:compute_analysis}
\end{figure}

\noindent\underline{\textbf{Computation Time.}}
In the XAI model, computation time primarily depends on sample size ($n$) and the total number of ROIs ($J$). To assess scalability, we fit XAI for $J=100, 300, 500, 1000$ and $n=300, 1000, 3000, 5000$, training each for $200$ epochs with batch size $64$ and using MC dropout with $F=200$ samples. As shown in Figure~\ref{fig:compute_analysis}, computation time scales linearly with both $n$ and $J$. For instance, full inference with 
$n=5000$ and $J=300$ takes under $5$ minutes.



\section{Application to the Multimodal ABCD Data}\label{sec:data_analysis}
\subsection{Data Structure and Methods}
We apply our proposed method to the ABCD dataset to answer our motivating scientific question described in Section \ref{sec:data_description}. Each participant is observed with two structural images containing 148 ROIs and these ROIs are mapped to a 3-dimensional space using 3-dimensional coordinates obtained from the R package \texttt{brainGraph} \citep{brainGraph}. Note that the raw outcome BOLDVAR contains some extreme values that fell well beyond the main density of the data. Therefore, we trimmed values greater than 3 which excluded most of these extreme values and resulted in missing data for 520 subjects yielding a total of 8655 subjects for the analysis data set. To facilitate interpretation of the targeted varying coefficient functions across ROIs, we scaled and centered both the log transformed outcome and local covariates within each ROI using the training data, such that they have a mean of zero and a standard deviation (SD) of one across subjects. This scaling yields standardized regression coefficients that are directly comparable across ROIs allowing us to compare effects in terms of SD units. To incorporate the network structure into our model, we employ a latent space model described in Section \ref{sec:methods}.2.1, which generates 3-dimensional latent variables representing the network. 
For the network nodes, we define each node as a sub-group of ROIs organized based on similar anatomical location defined by hemisphere and lobe as described in Section \ref{sec:data_description}. This hierarchical approach reduces the overall network complexity, allowing us to model the structure using only 12 nodes, despite the initial set containing 148 ROIs. This reduction in complexity is justified empirically based on preliminary analyses that considered co-activation networks among the full set of ROIs which produced comparable explanations of variation in the outcome but with much greater computational demands to perform the decomposition described in Section \ref{sec:methods}.2.1. Our analysis uses a total of $N=5,193$ participants for model fitting, $N_{\text{val}}=1,731$ participants for validation, and $N_{\text{test}}=1,731$ participants for testing. We utilized validation data to find out the optimized tunning hyper-parmaeters such as batch size, learning rate, and number of layers. We compare our inferential performance with competitors in terms of R-squared ($R^2$) values computed as the proportion of variation explained for frequentist methods and as in \citet{gelman2019r} for Bayesian procedures. Here, the batch size is 30 with a maximum of 500 epochs, and, as mentioned in the simulation settings, we apply the same early–stopping rule and learning–rate scheduling based on the validation mean squared error. The BART and BIRD–GP models are also trained following the same configurations used in the simulation study.

\subsection{Data Analysis Results} 

\begin{figure}[ht!]
    \centering
    \includegraphics[scale=0.45]{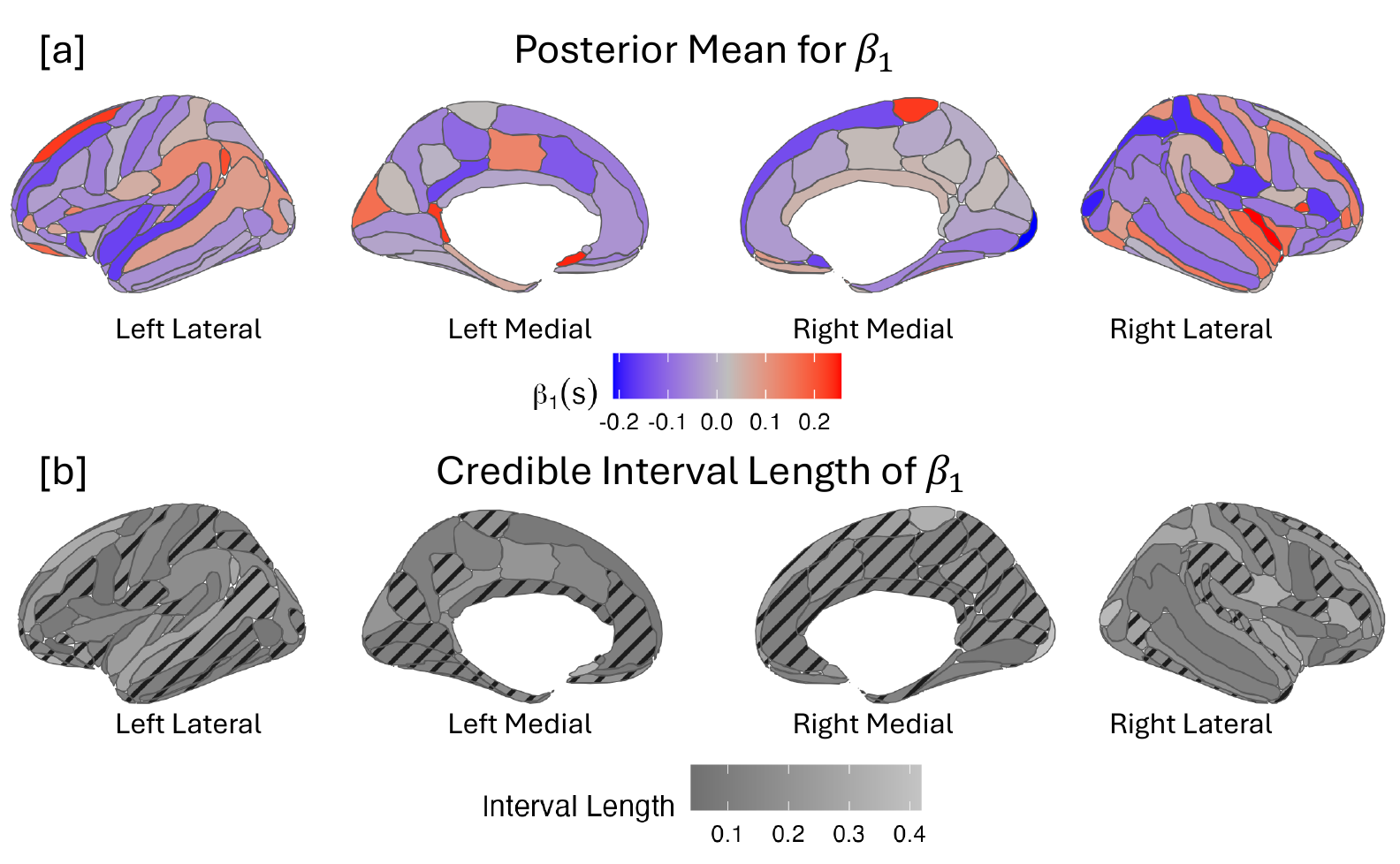} 
    \caption{Lateral and medial cross sections of the posterior \textbf{(a)} mean and \textbf{(b)} 95\% credible interval length for the spatially varying coefficient function $\bbeta_1(\bs_{v,j})$ capturing effects of GWMIC. When the posterior 95\% credible interval contains zero, the interval length is marked with back stripes in the figure as indication of uncertainty of any association.}
    \label{fig:beta1}
\end{figure}

We present the results from our application of the proposed XAI model to the multimodal ABCD imaging data. Figure \ref{fig:beta1} displays lateral and medial cross sections of the posterior (a) mean and (b) $95\%$ credible interval length for the spatially varying coefficient function $\bbeta_1(\bs_{v,j})$ capturing effects of GWMIC. In Figure \ref{fig:beta1}(a), positive (negative) regression coefficients can be interpreted as indicating a positive (negative) association between regions with greater cortical gray and white matter intensity contrasts and BOLDVAR. Figure  \ref{fig:beta1}(b) shows the length of the 95\% credible intervals (CI) for the estimated coefficient function, offering a measure of uncertainty—longer intervals reflect greater uncertainty, while shorter intervals indicate less uncertainty. When the posterior 95\% credible interval contains zero, the interval length is marked with back stripes in the Figure \ref{fig:beta1}(b). Roughly half of the 95\% credible intervals for ROI-effects exclude zero (80/148 ROIs, 54.1\%), where the greatest proportion of CIs excluding zero occurs in the insula (12/24 ROIs, 75.0\%) and frontal (27/44 ROIs, 61.4\%) suggesting these lobes are where our XAI model is most certain of an association between GWMIC and the outcome. Focusing on the coefficients with 95\% CIs that exclude zero within each lobe, estimated effects range from $-0.22$ to $0.26$, indicating that within each ROI a SD increase in GWMIC is associated with generally less than a $0.26$ SD shift in the outcome log(BOLDVAR) reflecting a small standardized effect size. The estimated coefficient function shows a mixture of positive and negative associations within each lobe indicating a variable pattern of associations between GWMIC and BOLDVAR. Situating these results in relation to prior research, \citet{Zuo2010} found that gray matter in the brain displays higher ALFF than white matter, though these results focus on gray and white matter parcellations of the brain as opposed to cortical contrasts analyzed here. In fact, the relationship between BOLDVAR and GWMIC is essentially unexplored in adolescent children and these findings suggest a mixture of associations both within and across brain lobes when controlling for the contributions of CT and multi-task coactivation. 

\begin{figure}[ht!]
    \centering
    \includegraphics[scale=0.45]{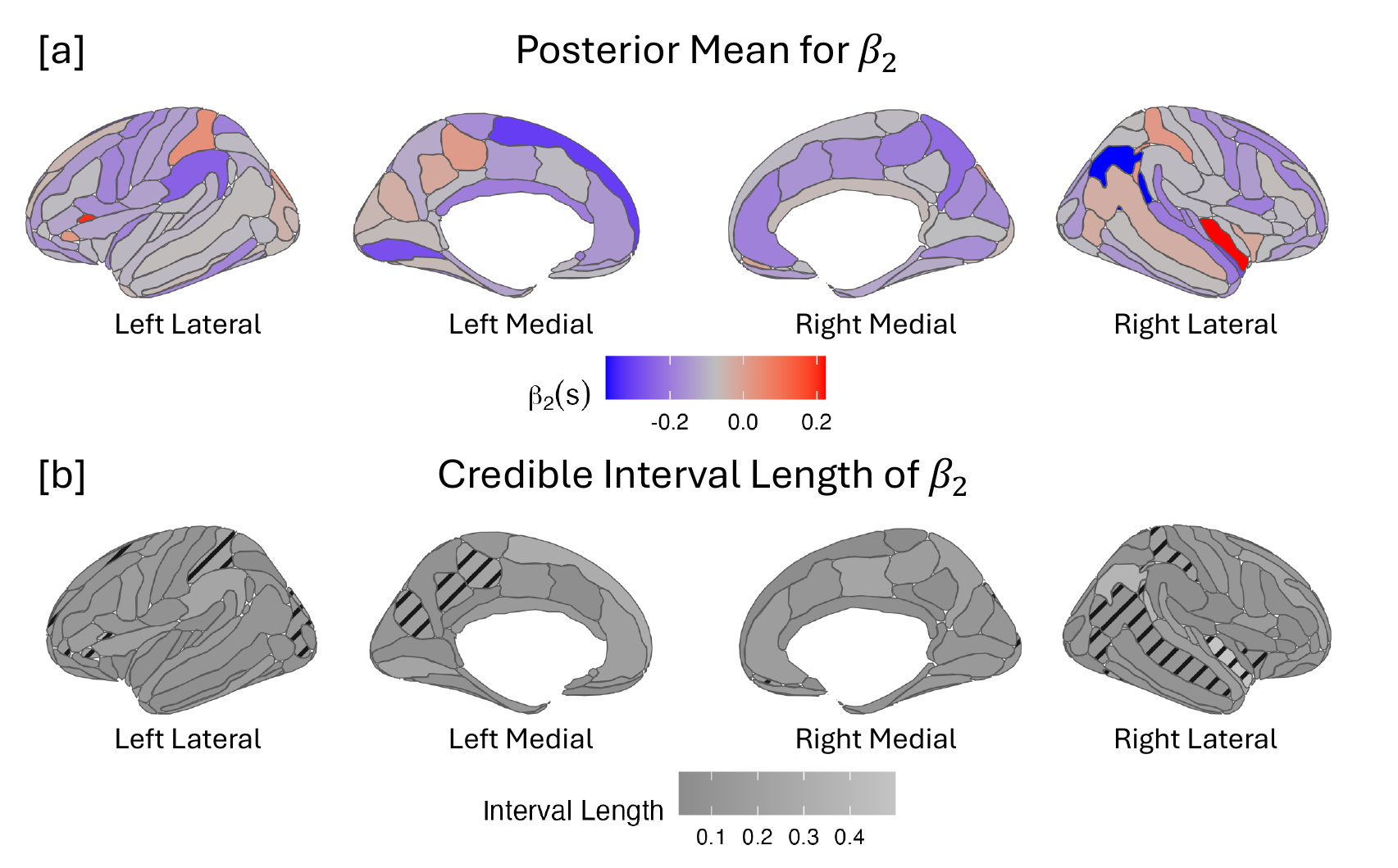} 
    \caption{Lateral and medial cross sections of the posterior \textbf{(a)} mean and \textbf{(b)} 95\% credible interval length for the spatially varying coefficient function $\bbeta_2(\bs_{v,j})$ capturing effects of CT. When the posterior 95\% credible interval contains zero, the interval length is marked with back stripes in the figure as indication of uncertainty of any association.}
    \label{fig:beta2}
\end{figure}

Figure \ref{fig:beta2} displays lateral and medial cross sections of the posterior (a) mean and (b) $95\%$ credible interval length for the spatially varying coefficient function $\bbeta_2(\bs_{v,j})$ capturing effects of CT. Figure \ref{fig:beta2}(b), displays the $95\%$ CI length for the estimated coefficient function and are interpreted as described above. Generally, 95\% CIs for ROI-effects exclude zero (124/148 ROIs, 83.8\%), suggesting our XAI model is relatively certain of the association between CT even in the occipital lobe (21/28 ROIs, 75.0\%) which displays the greatest uncertainty of non-zero effects for CT. Focusing on the coefficients with 95\% CIs that exclude zero within each lobe, estimated effects range from $-0.38$ to $-0.05$, indicating that within each ROI a SD increase in CT is associated with up to a $0.38$ SD decrease in the outcome log(BOLDVAR) reflecting a small to medium standardized effect size. The estimated coefficient function shows exclusively negative associations with the strongest signals in the temporal and limbic lobes. There is essentially no prior research on the relationship between BOLDVAR and CT in adolescent children and these results suggest a near uniform negative association throughout the brain that varies in magnitude from no effect to a medium effect size when controlling for the contributions of GWMIC and multi-task coactivation.

\begin{table} [h!]
\centering
\caption[]{Predictive inference results for real data application from different methods. For all methods, overall and average $R^2$, prediction coverage, 95\% HPD interval, and computing time (min) are reported in the table. \dag  Frequentist $R^2$ is reported for the GLM models.}
\label{table_result}
\begin{tabular}{l l c c c c c c c}
  & &  \textbf{XAI} &\textbf{BART} & \textbf{BIRD-GP} &\textbf{GLM} \\ 
 &  &  & & & & \\
\hline\hline
 R-squared (overall) &  & 0.029 & - & 0.011  & 0.026 \dag \\
  & Coverage  & 0.942 & 0.088 & 0.930 & -  \\
 & Intervals  & 3.804  & 0.222 &  3.627 & -  \\
 Time (min) && 29.763 &  130.421 & 170.745  & 0.125  \\
\hline
\hline
\end{tabular}
\label{tab:real_data}
\end{table}

Table \ref{tab:real_data} displays predictive performance and computational efficiency metrics for all methods, including R-squared, coverage, and interval width for our technique compared to the listed competitors. We report an overall $R^2$ computed by aggregating predictions across all ROIs without region-wise separation. In terms of our ABCD data, the $R^2$ suggests that our proposed XAI model explains the outcome image better than more complex models such as BIRD-GP, which allow for nonlinear interactions between imaging modalities. While BART and GLM cannot provide posterior predictive means and thus its Bayesian $R^2$ is not available. The frequentist $R^2$ for GLM is available and in the same range as our proposed method providing a qualitative comparison. As shown in Supplementary Figure 3, there is structural heterogeneity in $R^2$ performance, with some regions being well explained and others poorly captured, which demonstrates the region-specific differences among ROIs. Summarizing the results by lobe, the $R^2$ is highest in the temporal lobe followed by the frontal and occipital lobes, with the lowest predictive performance in the limbic lobe. We compare the $R^2$ for our XAI model with and without the network-valued inputs. Section 3 of the supplementary file shows a small improvement by including network-valued predictors compared to a model with structural information alone.

Our proposed XAI model provides nominal coverage followed by BIRD-GP with near nominal coverage and BART with severe under coverage (see Table \ref{tab:real_data}). In terms of credible interval length, BART provides the shortest intervals, explaining its poor coverage. In terms of computation, the XAI takes approximately four times less than other competitors that offer posterior predictive inference. In conclusion, XAI provides equivalent point prediction while yielding superior predictive uncertainties and much faster computation which are of critical importance when dealing with large neuroimaging data sets.

\section{Conclusion}
This article proposes a framework for estimating spatially-indexed spatially indexed low frequency fluctuations (ALFF) in resting state functional MRI as a function of cortical structural features and a multi-task coactivation network. The framework leverages deep neural networks to model complex nonlinear relationships between the input spatial and network-valued images and the spatial output image. It enables accurate estimation of the effect of each input image on the outcome image while providing crucial uncertainty quantification within DNN—both for prediction and for effect estimation—through a MC dropout strategy. In simulations and a real data analysis, the proposed explainable AI (XAI) approach demonstrates excellent performance in point estimation and uncertainty quantification. Additionally, the model scales linearly with the sample size and the number of brain regions, ensuring computational efficiency in high-resolution and large-scale neuroimaging studies.

While this article assumes a common relationship between the activation map, cortical metrics, and brain connectivity network across all subjects, an important direction for future work is to investigate how these relationships vary across individuals. This motivates the development of model-based clustering strategies, wherein the proposed XAI framework is fitted within each identified cluster. This extension is the focus of our ongoing research.

\section{Acknowledgment }
Dr. Rajarshi Guhaniyogi is supported by National Science Foundation DMS-2210672, and National Institute of Health R01NS131604.

\bibliography{References}

\end{document}


\title{Supplementary material of Deep Generative Modeling with Spatial and Network Images: An Explainable AI (XAI) Approach}






\maketitle

\section{Details of Deep Gaussian Process Approximation of Deep Neural network with Dropout}
The parameter $\bDelta$ of the variational distribution described in Equation (10) of the main article is optimized by maximizing $\mbox{E}_{\tilde{q}}[{\log(\pi(\boldsymbol{\theta},\mathbf{D}))}]-\mbox{E}_{\tilde{q}}[\log \tilde{q}({\boldsymbol{\theta}}|\bDelta)]$, the evidence lower bound (ELBO). Denote the full prior distribution on the parameters $\btheta$ by
$p(\btheta)=\prod_{l_1=1}^{L_\beta}\prod_{q=1}^Q p(\mathbf{W}^{(\beta)}_{q,l_1})p(\mathbf{b}^{(\beta)}_{q,l_1})\prod_{l_2=1}^{L_h}p(\mathbf{W}^{(h)}_{l_2})p(\mathbf{b}^{(h)}_{l_2})$. The log ELBO is then given by,
\begin{align}\label{gaussianVI}
   & \mathcal{L}_{\text{GP-VI}}(\tau^2,\bDelta) =  \sum_{i=1}^{n}\sum_{v=1}^{V}\sum_{j=1}^{J_v}\int \cdots \int \tilde{q}(\btheta|\bDelta) \log p(y_{i}(\bs_{v,j})|\bx_{i}(\bs_{v,j}),\bg_{i,v},\tau^2, \{\boldsymbol{\Psi}^{(\beta)}_{q,L_\beta}\}_{q=1}^Q,\boldsymbol{\Psi}^{(h)}_{L_h})\nonumber\\
     &  \prod_{l_1=2}^{L_\beta}\prod_{q=1}^Q p(\boldsymbol{\Psi}^{(\beta)}_{q,l_1}|\boldsymbol{\Psi}^{(\beta)}_{q,l_1-1},\bW_{q,l_1}^{(\beta)},\bb_{q,l_1}^{(\beta)}) 
     \prod_{l_2=2}^{L_h} p(\boldsymbol{\Psi}^{(h)}_{l_2}|\boldsymbol{\Psi}^{(h)}_{l_2-1},\bW_{l_2}^{(h)},\bb_{l_2}^{(h)})d\mathbf{W}^{(\beta)}_{q,l_1} d\mathbf{b}^{(\beta)}_{q,l_1} d\mathbf{W}^{(h)}_{l_2} d\mathbf{b}^{(h)}_{l_2}
     \quad  \nonumber\\
     & - \text{KL}\Big (\tilde{q}(\btheta|\bDelta) \Big|\Big | p(\btheta) \Big ),
\end{align}
where $p(\mathbf{y}_{i}(\bs_{v,j})|\bx_{i}(\bs_{v,j}), \bg_{i,v},\tau^2,\boldsymbol{\Psi}^{(\beta)}_{q,L_\beta},\boldsymbol{\Psi}^{(h)}_{L_h})$ is the data likelihood, obtained from Equation (9) in the main article. Note that $\mathbf{\bPsi}^{(\beta)}_{q,L_\beta}$ and $\mathbf{\bPsi}^{(h)}_{L_h}$ are dependent on the weight and bias parameters $\lbrace \mathbf{W}^{(\beta)}_{q,l_1},\mathbf{b}^{(\beta)}_{l_1}\rbrace_{l_1=1}^{L_\beta}$ and $\lbrace \mathbf{W}^{(h)}_{l_2},\mathbf{b}^{(h)}_{l_2}\rbrace_{l_2=1}^{L_h}$, respectively.

Let $\bz_{q,l_1}^{(\beta)(f)}\in \{ 0,1\}^{k_{l_1}^{(\beta)}}$ be a dropout mask with independent components distributed as $\text{Bernoulli}(1-p^{(\beta)}_{l_1})$, corresponding to the  
$l_1$th layer ($l_1 = 1, \dots, L_\beta-1$) of the network used to model the function $\beta_q(\cdot)$, and indexed by iteration $f = 1, \dots, F$. Likewise, let $\bz_{l_2}^{(h)(f)}\sim \text{Bernoulli}(1-p^{(h)}_{l_2})$ denote the dropout mask of dimension $k_{l_2}^{(h)}$ associated with the $l_2$th layer ($l_2=1,\cdots,L_h-1)$ in the modeling the function $h(\cdot)$. We define a matrix $\bZ_{q,l_1}^{(\beta)(f)}=\bz_{q, l_1}^{(\beta)(f)} {\boldsymbol 1}_{k_{l_1-1}^{(\beta)}}$, where ${\boldsymbol 1}_{k_{l_1-1}^{(\beta)}}$ is a vector of ones. Analogous constructions hold for the modeling of $h(\cdot)$ function, with $\bZ_{l_2}^{(h)(f)} = \bz_{l_2}^{(h)(f)} {\boldsymbol 1}_{k_{l_2-1}^{(h)}}$.

For all layers and outcome for $\beta_q(\cdot)$ and $h(\cdot)$, the complete collections of dropout masks at iteration $f$ are defined as  
\begin{align*}
  \bz_{q}^{(\beta)(f)} &= \left\{ \left( \operatorname{vec}(\bZ_{q,l_1}^{(\beta)(f)})^\top,\, (\bz_{q,l_1}^{(\beta)(f)})^\top \right)^\top :\, l_1 = 1, \ldots, L_\beta - 1\right\} \\  
   \bz^{(h)(f)} &= \left\{ \left( \operatorname{vec}(\bZ_{l_2}^{(h)(f)})^\top,\, (\bz_{l_2}^{(h)(f)})^\top \right)^\top :\, l_2 = 1, \ldots, L_h - 1\right\}. \\ 
\end{align*}
The full dropout collection for iteration $f$ is defined as 
\[\bz^{(m)} = \left( (\bz_q^{(\beta)(f)})^\top,\, (\bz^{(h),(f)})^\top \right)^\top.\]
For each iteration $f$, let $\bepsilon^{(f)}$ denote a Gaussian perturbation vector, independent of the dropout, with $\bepsilon^{(f)} \sim N(\boldsymbol{0},\, \sigma^2 \bI)$, and of the same dimension as the parameter vector $\btheta$. Under the variational distribution in Equation (10) of the main article, with variational mean vector $\bDelta$, each sample in iteration $f$ is generated as 
\[
\btheta^{(f)} = \bDelta \odot \bz^{(f)} + \bepsilon^{(f)},
\]
where $\odot$ denotes the element-wise product. Consequently, the variational samples $\btheta^{(f)}$ can be viewed as independent draws from the joint distribution of $(\bz^{(f)\top},\, \bepsilon^{(f)\top})^\top.$ 

The variational ELBO can be written as
\[
\mathcal{L}_{\mathrm{GP-VI}}(\tau^2,\bDelta) = \mathbb{E}_{q(\btheta \mid \bDelta)}\left[ \log p(\btheta, \bD) \right] - \mathbb{E}_{q(\btheta \mid \bDelta)} \left[ \log q(\btheta \mid \bDelta) \right].
\]
This can be rewritten explicitly in terms of the auxiliary variables:
\begin{align}
\mathcal{L}_{\mathrm{GP-VI}}(\bDelta, \tau^2)
&= \mathbb{E}_{q(\btheta \mid \bDelta)} \left[ \log p(\bD \mid \btheta) \right] - \mathrm{KL} \left( q(\btheta \mid \bDelta) \,\|\, p(\btheta) \right) \nonumber \\
&= \mathbb{E}_{\bz,\, \bepsilon} \left[ \log p(\bD \mid \bDelta,\, \bz,\, \bepsilon) \right] - \mathrm{KL} \left( q(\btheta \mid \bDelta)\,\|\,p(\btheta) \right) \nonumber \\
&= \int \log p(\bD \mid \bDelta,\, \bz,\, \bepsilon) \, dP_{\bz,\, \bepsilon}(\bz,\, \bepsilon) - \mathrm{KL}\left( q(\btheta \mid \bDelta)\,\|\,p(\btheta) \right).
\end{align}
For numerically approximate the expectation $\int \log p(\bD \mid \bDelta,\, \bz,\, \bepsilon) \, dP_{\bz,\, \bepsilon}(\bz,\, \bepsilon)$, we draw $F$ independent samples $\lbrace(\bz^{(f)},\bepsilon^{(f)})\rbrace_{f=1}^{F}$ and use the Monte Carlo estimator: 
\[
\int \log p(\bD \mid \bDelta,\, \bz,\, \bepsilon) \, dP_{\bz,\, \bepsilon}(\bz,\, \bepsilon)
\; \approx \;
\frac{1}{F} \sum_{f=1}^{F} \log p(\bD \mid \btheta^{(f)}),
\]
where $\btheta^{(f)} = \bet \odot \bz^{(f)} + \bepsilon^{(f)}$. When the variance of the Gaussian perturbation is negligible ($\sigma^2 \approx 0$), the samples satisfy $\btheta^{(f)} \approx \bDelta \odot \bz^{(f)},$ which is often sufficient in practice.

Hence, we write the likelihood $p(y_{i}(\bs_{v,j})| \mathbf{x}_i(\bs_{v,j}),\bg_{i,v},\tau^2, \mathbf{\Psi}^{(\beta)}_{q,L_\beta},\mathbf{\Psi}^{(h)}_{L_h})$ more compactly as $p(y_{i}(\bs_{v,j})| \mathbf{x}_i(\bs_{v,j}),\bg_{i,v},\tau^2, \btheta)$. The conditional distributions $p(\boldsymbol{\Psi}^{(\beta)}_{q,l_1}|\boldsymbol{\Psi}^{(\beta)}_{q,l_1-1},\bW_{q,l_1}^{(\beta)},\bb_{q,l_1}^{(\beta)})$ and $p(\boldsymbol{\Psi}^{(h)}_{l_2}|\boldsymbol{\Psi}^{(h)}_{l_2-1},\bW_{l_2}^{(h)},\bb_{l_2}^{(h)})$ are Gaussian likelihoods obtained from Equation (2) of the main article. Substituting them into the variational objective yields the MC approximation of the log ELBO
\begin{equation}
\begin{split}
&\mathcal{L}_{\text{GP-MC}}(\tau^2,\bDelta) =\frac{1}{F}\sum_{f=1}^{F}\sum_{i=1}^{n} \sum_{v=1}^{V}\sum_{j=1}^{J_v} \log p(\mathbf{y}_{i}(\bs_{v,j})|\bx_{i}(\bs_{v,j}),\bg_{i,v},\tau^2,\btheta^{(f)}) \\
&\qquad\qquad\qquad\qquad- \text{KL}\Big (\tilde{q}(\btheta|\bDelta) \Big|\Big | p(\btheta) \Big ),
\label{GPMC}
\end{split}
\end{equation} 
where $\btheta^{(m)}=\{\lbrace \mathbf{W}_{q,l_1}^{(\beta)(m)}, \mathbf{b}_{q,l_1}^{(\beta)(m)}\rbrace_{l_1=1,q=1}^{L_\beta, Q}$, $\lbrace \mathbf{W}_{l_2}^{(h)(m)}, \mathbf{b}_{l_2}^{(h)(m)} \rbrace_{l_2=1}^{L_h}\}$ are MC samples from the variational distribution given in Equation (10) of the main article. 

As demonstrated in \citet{gal2016dropout}[Proposition 1], under the conditions where the number of neurons $k_{l_1}^{(\beta)}$ and $k_{l_2}^{(h)}$ are large, and $\sigma$ in Equation (10) is small, the KL divergence between $\tilde{q}(\btheta|\bDelta)$ and $p(\btheta)$ can be approximated as
\begin{equation}\begin{split}
 & \sum_{l_{1}=1}^{L_\beta}\sum_{q=1}^{Q}\frac{p^{(\beta)}_{l_1}}{2}(||\bmu_{q,l_{1}}^{w,(\beta)}||_2^2)+Q\sum_{l_1=1}^{L_\beta}k_{l_1}^{(\beta)}(\sigma^{2} -(1+\log2\pi)-\log\sigma^{2}+C_1) \\ 
 &+ \sum_{l_{1}=1}^{L_\beta}\sum_{q=1}^{Q}\frac{p^{(\beta)}_{l_1}}{2}(||\bmu_{q,l_{1}}^{b,(\beta)}||_2^2)+Q\sum_{l_1=1}^{L_\beta}k_{l_1}^{(\beta)}(\sigma^{2} -(1+\log2\pi)-\log\sigma^{2}+C_2) \\ 
 &+\sum_{l_{2}=1}^{L_h}\frac{p^{(h)}_{l_2}}{2}(||\bmu_{l_{2}}^{w,(h)}||_2^2)+\sum_{l_2=1}^{L_h}k_{l_2}^{(h)}(\sigma^{2} -(1+\log2\pi)-\log\sigma^{2}+C_3) \\ 
 &+ \sum_{l_{2}=1}^{L_h}\frac{p^{(h)}_{l_2}}{2}(||\bmu_{l_{2}}^{b,(h)}||_2^2)+\sum_{l_2=1}^{L_h}k_{l_2}^{(h)}(\sigma^{2} -(1+\log2\pi)-\log\sigma^{2}+C_4), \\ 
\end{split}
\label{KLdiver}
\end{equation}
with constants $C_1,C_2,C_3,C_4$ involving $k_{l_1}^{(\beta)},k_{l_2}^{(h)}$ ($l_1=1,..,L_\beta$; $l_2=1,..,L_h$) and $\sigma^2$. Here $\bmu_{q,l_1}^{w,(\beta)}=(\mu_{q,l_1,kk'}^{w,(\beta)}:k,k'=1,
\ldots,k_{l_1}^{(\beta)})^T$, $\bmu_{l_2}^{w,(h)}=(\mu_{l_2,kk'}^{w,(h)}:k,k'=1,..,k_{l_2}^{(h)})^T$, $\bmu_{q,l_1}^{b,(\beta)}=(\mu_{q,l_1,k}^{b,(\beta)}:k=1,\ldots,k_{l_1}^{(\beta)})^T$ and $\bmu_{l_2}^{b,(h)}=(\mu_{l_2,k}^{b,(h)}:k=1,\ldots,k_{l_2}^{(h)})^T$ denote the variational mean vectors associated with the weights and biases of the corresponding layers.

Plugging $\text{KL}\Big (\tilde{q}(\btheta|\bDelta) \Big|\Big | p(\btheta) \Big ) $ approximation to \eqref{GPMC}, \begin{align}\label{GPMCKLsuppl}
&\mathcal{L}_{\text{GP-MC}}(\tau^2,\bDelta) \approx \frac{1}{M}\sum_{m=1}^{M}\sum_{i=1}^{n} \sum_{v=1}^{V}\sum_{j=1}^{J_v} \log p(\mathbf{y}_{i}(\bs_{v,j})|\bx_{i}(\bs_{v,j}),\bg_{i,v},
\tau^2,\btheta^{(m)})-\sum_{l_{1}=1}^{L_\beta}\sum_{q=1}^{Q}\frac{p^{(\beta)}_{l_1}}{2}(||\bmu_{q,l_{1}}^{w,(\beta)}||_2^2) \nonumber \\
&\qquad\qquad\qquad-\sum_{l_{1}=1}^{L_\beta}\sum_{q=1}^{Q}\frac{p^{(\beta)}_{l_1}}{2}(||\bmu_{q,l_{1}}^{b,(\beta)}||_2^2)-\sum_{l_{2}=1}^{L_h}\frac{p^{(h)}_{l_2}}{2}(||\bmu_{l_{2}}^{w,(h)}||_2^2)-\sum_{l_{2}=1}^{L_h}\frac{p^{(h)}_{l_2}}{2}(||\bmu_{l_{2}}^{b,(h)}||_2^2)\nonumber\\
&-Q\sum_{l_1=1}^{L_\beta}k_{l_1}^{(\beta)}(\sigma^{2} -(1+\log2\pi)-\log\sigma^{2}+C_1)-Q\sum_{l_1=1}^{L_\beta}k_{l_1}^{(\beta)}(\sigma^{2} -(1+\log2\pi)-\log\sigma^{2}+C_2)\nonumber\\
&-\sum_{l_2=1}^{L_h}k_{l_2}^{(h)}(\sigma^{2} -(1+\log2\pi)-\log\sigma^{2}+C_3)-\sum_{l_2=1}^{L_h}k_{l_2}^{(h)}(\sigma^{2} -(1+\log2\pi)-\log\sigma^{2}+C_4)
\end{align}
By ignoring constant hyperparameter $\sigma$, the approximated version of the ELBO scaled by a positive constant $\frac{1}{N}$ becomes  
\begin{align} 
     &\mathcal{L}_{\text{GP-MC}}(\tau^2,\bDelta) \approx -\frac{1}{N}\sum_{i=1}^{n}\sum_{v=1}^{V}\sum_{j=1}^{J} ||{\by}_{i,v}(\bs_{v,j}) -\widehat{\by}_{i,v}(\bs_{v,j})||_2^2 - \sum_{l_1=1}^{L_\beta}\sum_{q=1}^Q\frac{p^{(\beta)}_{l_1}}{2 N} ||\boldsymbol{\mu}^{w,(\beta)}_{q,l_1}||_2^2 \nonumber\\ 
     &\qquad\qquad - \sum_{l_1=1}^{L_\beta}\sum_{q=1}^Q\frac{p^{(\beta)}_{l_1}}{2 N}||\boldsymbol{\mu}^{b,(\beta)}_{q,l_1}||_2^2  - \sum_{l_2=1}^{L_h}\frac{p^{(h)}_{l_2}}{2 N} ||\boldsymbol{\mu}^{w,(h)}_{l_2}||_2^2 - \sum_{l_2=1}^{L_h}\frac{p^{(h)}_{l_2}}{2 N}||\boldsymbol{\mu}^{b,(h)}_{l_2}||_2^2.
\label{GPMC1}
\end{align}

\section{Impact of Changing Correlation Parameters in Simulation Study}

We employ the identical data generation strategy described in Section 4 of the main article. The main article already shows performance of XAI and competitors in Cases 1-7 by varying the number of ROIs $J$ and by varying SNRs.
To assess the impact of varying spatial correlation, we consider three additional cases by varying the spatial scale parameters $\eta_{\beta,1}$, $\eta_{\beta,2}$, and $\eta_h$ (see Supplementary Table~\ref{changecorel}) while keeping the variance parameters fixed. Increasing the scale parameters induces stronger association, allowing us to evaluate how model performance changes with the degree of association in $\beta_1(\cdot)$, $\beta_2(\cdot)$, and $h(\cdot)$. Here, we use a batch size of 10 and train for up to 500 epochs. We apply the same early–stopping rule based on the validation mean squared error. For the competitor models, BIRD-GP is implemented with 30 basis functions and optimized using 2,500 gradient steps, and BART is trained using an ensemble of 200 trees.

\begin{figure}[h!]
\centering\includegraphics[scale=0.7]{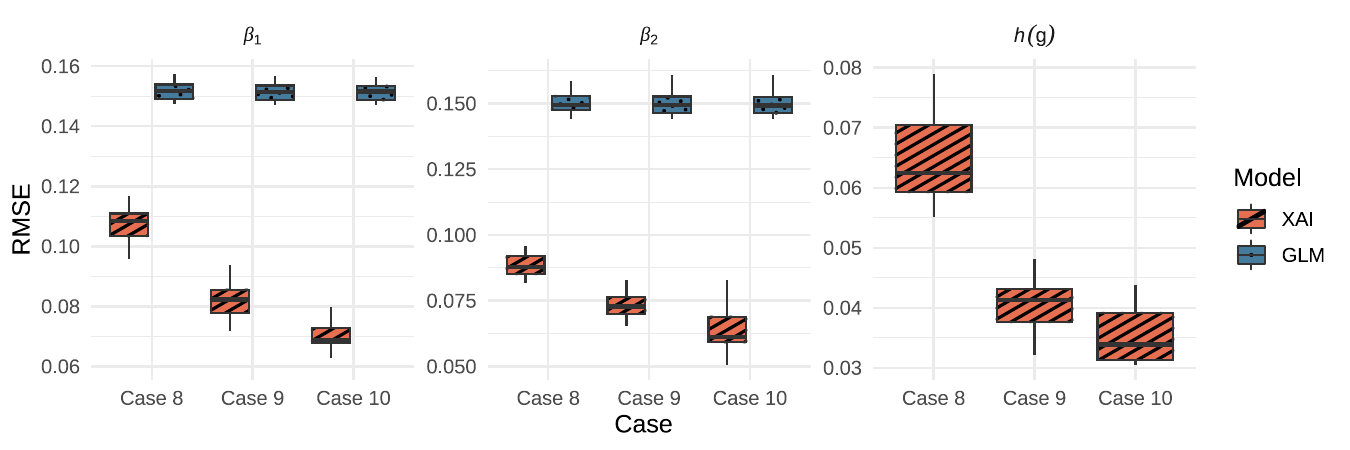}
    \caption{Root mean squared error (RMSE) comparison between the proposed XAI method and GLM for estimating the spatially varying coefficients $\beta_1(\bs)$ and $\beta_2(\bs)$, and the function encoding network latent effect $h(\bg)$ under Case 8 to Case 10. Boxplots show the distribution of RMSE across $100$ simulation replications. The inference on $\beta_1(\bs),\beta_2(\bs),h(\bg)$ are not available for BART and BIRD-GP. The inference on $h(\bg)$ is not available for GLM.}
\label{fig:CaseC_betas}
\end{figure}
\begin{figure}[h!]
\centering\includegraphics[scale=0.6]{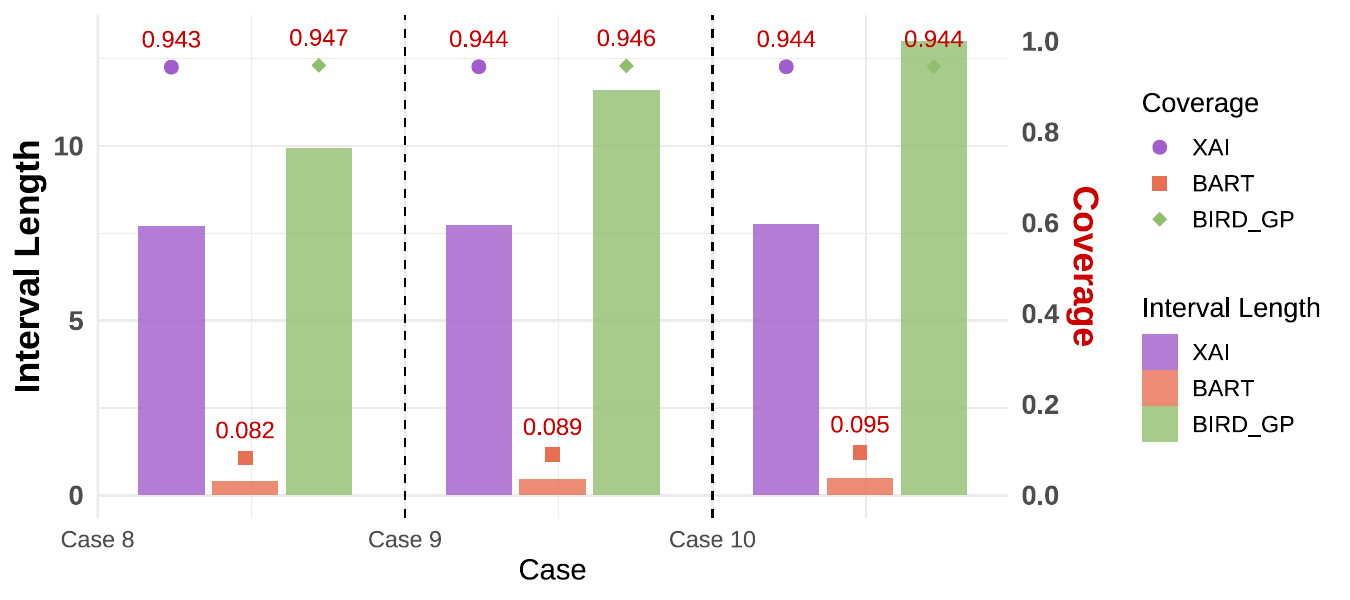}
    \caption{Coverage and length of 95\% predictive intervals averaged over replications for XAI, BIRD-GP and BART. They are not available for GLM. Results show near nominal coverage of XAI with smaller predictive intervals than BIRD-GP.}
\label{fig:CaseC}
\end{figure}

An increase in the long-range dependence from Cases 8-10 becomes less favorable to GLM that does not incorporate dependence, leading to an increasing performance gap between GLM and XAI in terms of estimating spatially varying coefficients $\beta_1(\bs),\beta_2(\bs)$ and the function $h(\bg)$, as evident in Supplementary Figure~\ref{fig:CaseC_betas}. As shown in Supplementary Table~\ref{changecorel} and Supplementary Figure~\ref{fig:CaseC}, XAI outperforms BIRD-GP in point prediction, and allows narrower predictive intervals with a similar coverage. As before, BART performs comparably in terms of point prediction but underestimates uncertainty.

\begin{table}[h]
\centering
\caption[]{Prediction performance comparison for simulated datasets  by varying the scale parameters $\eta_{\beta,1},\eta_{\beta,2},\eta_h$, while fixing the variance parameters. The table reports the root mean squared prediction error (RMSPE) with standard errors in parentheses, for each method (XAI, BART, BIRD-GP, and GLM). The standard errors are computed over $100$ replications.}\label{changecorel}
\begin{tabular}{ccclcccc}
$\eta_{\beta,1}$ & $\eta_{\beta,2}$ & $\eta_h$ & Case  & XAI & BART & BIRD-GP & GLM \\
\hline\hline
2 & 3 & 3  & Case 8 & \textbf{2.0048(0.005)} & 2.0150(0.005)  & 2.6311(0.391)  & 2.0339(0.007)  \\
5 & 6 & 5   & Case 9 & \textbf{2.0025(0.006)} & 2.0086(0.005)  & 2.6171(0.400)  & 2.0267(0.005)  \\
7 & 8 & 8  & Case 10 & \textbf{2.0017(0.006)}  &  2.0072(0.005) & 2.6136(0.409)  &  2.0259(0.004) \\
\hline\hline
\end{tabular}
\end{table}

\section{Region-Wise $R^2$ Comparison}
\begin{figure}
\centering
   \includegraphics[scale=0.46]{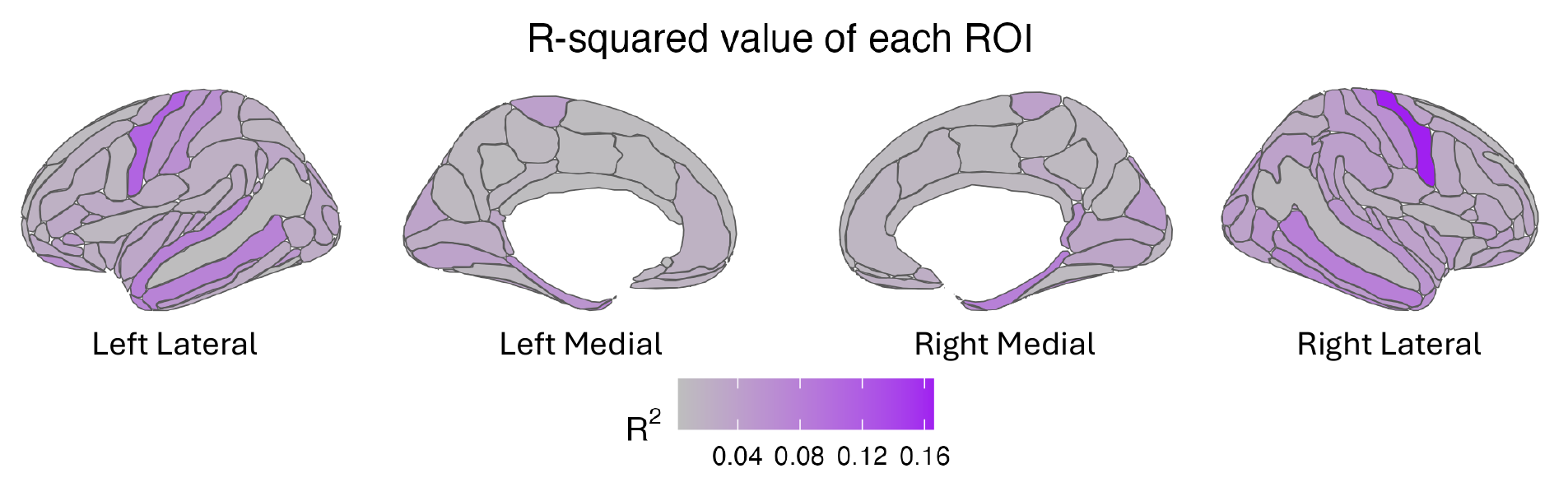}
   \caption{$R^2$ values for each ROI. Higher $R^2$ values (in darker purple) indicate ROIs where the model provides better explanatory power.}
\label{fig:R2_one}
\end{figure}

Supplementary Figure~\ref{fig:R2_one} shows that there is structural heterogeneity in $R^2$ performance, with some regions being well explained and others poorly captured, which demonstrates the region-specific differences among ROIs. Summarizing the results by lobe, the $R^2$ is highest in the temporal lobe followed by the frontal and occipital lobes, with the lowest predictive performance in the limbic lobe. 

\begin{figure}[htpb]
\centering
   \includegraphics[scale=0.35]{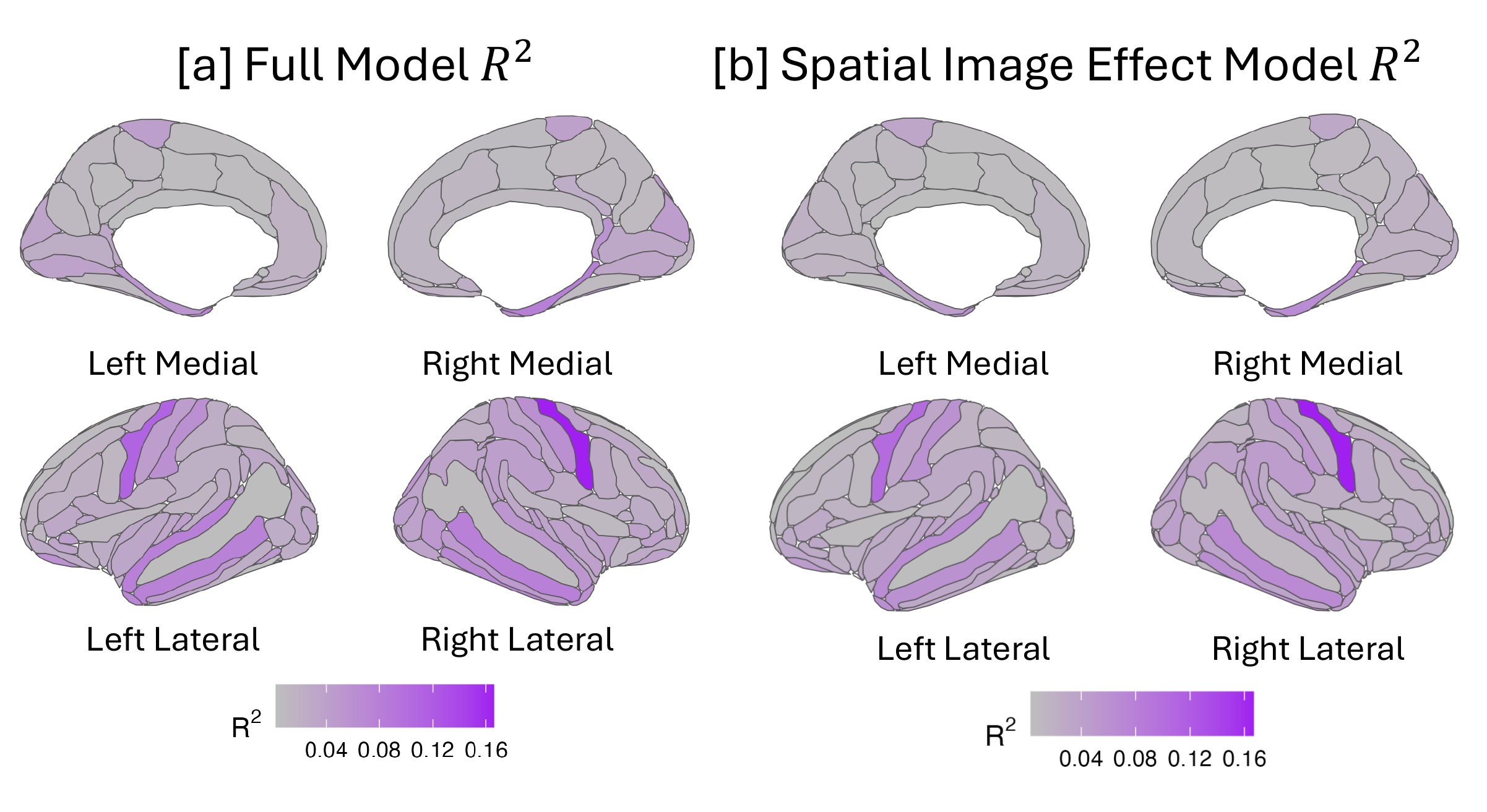}
   \caption{ROI-wise $R^2$ values from (a) the full model incorporating both spatial image effects and network-valued image effects, and (b) the model using only spatial image effects. Higher $R^2$ values (in darker purple or darker yellow) indicate brain regions where the model achieves greater explanatory power.}
\label{fig:R2_effects}
\end{figure}

We also compare the $R^2$ for our XAI model with and without the network-valued inputs. Supplementary Figure~\ref{fig:R2_effects} displays the distribution of ROI-wise $R^2$ values under two model settings: the left panel shows results from the full model incorporating both spatial image effects and network-valued image effects, while the right panel illustrates the model using only spatial image effects. When only spatial image effects are included, the average $R^2$ is 0.024. Upon integrating network-valued features, the average $R^2$ increases to 0.029, indicating that the network information provides some additional explanatory power.

\section{ABCD Data Acknowledgement}

Data used in the preparation of this article were obtained from the Adolescent Brain Cognitive DevelopmentSM (ABCD) Study (https://abcdstudy.org), held in the NIMH Data Archive (NDA). This is a multisite, longitudinal study designed to recruit more than 10,000 children age 9-10 and follow them over 10 years into early adulthood. The ABCD Study is supported by the National Institutes of Health and additional federal partners under award numbers U01DA041048, U01DA050989, U01DA051016, U01DA041022, U01DA051018,\\ 
U01DA051037, U01DA050987, U01DA041174, U01DA041106, U01DA041117, U01DA041028, U01DA041134, U01DA050988, U01DA051039, U01DA041156, U01DA041025, U01DA041120, U01DA051038, U01DA041148, U01DA041093, U01DA041089, U24DA041123, U24DA041147. A full list of supporters is available at \url{https://abcdstudy.org/federal-partners.html}. A listing of participating sites and a complete listing of the study investigators can be found at \url{https://abcdstudy.org/consortium_members}. ABCD consortium investigators designed and implemented the study and/or provided data but did not necessarily participate in the analysis or writing of this report. This manuscript reflects the views of the authors and may not reflect the opinions or views of the NIH or ABCD consortium investigators. The ABCD data repository grows and changes over time. The ABCD data used in this report came from doi: 10.15154/8873-zj65.

\bibliography{References}